\documentclass[aps,prx,twocolumn,longbibliography,superscriptaddress,noeprint,reprint]{revtex4-1}

\usepackage{graphicx}
\usepackage{mathtools}
\usepackage{tikz}
\usepackage{siunitx}
\usepackage{xcolor}
\usepackage[normalem]{ulem}
\usepackage{hyperref}

\usepackage{multirow}

\usepackage[english]{babel}

\usepackage{float}

\usepackage{sidecap}

\usepackage{pgffor}
\usepackage{pdfpages} 
\makeatletter
\AtBeginDocument{\let\LS@rot\@undefined}
\makeatother

\usepackage{xcolor}
\hypersetup{
    colorlinks,
    linkcolor={red!50!black},
    citecolor={blue!50!black},
    urlcolor={blue!80!black}
}
\usepackage{outlines}

\usepackage{color}
\definecolor{blue}{rgb}{0.00,0.00,0.95}

\AtBeginDocument{%
    \newwrite\bibnotes
    \def\bibnotesext{Notes.bib}
    \immediate\openout\bibnotes=\jobname\bibnotesext
    \immediate\write\bibnotes{@CONTROL{REVTEX41Control}}
    \immediate\write\bibnotes{@CONTROL{%
    apsrev41Control,author="08",editor="1",pages="1",title="0",year="1"}}
     \if@filesw
     \immediate\write\@auxout{\string\citation{apsrev41Control}}%
    \fi
}%

\begin{document}

\title{Information content and optimization of self-organized developmental systems}
\author{David B. Br\"uckner}
\email[]{david.brueckner@ist.ac.at}
\affiliation{Institute of Science and Technology, Am Campus 1, 3400 Klosterneuburg, Austria}
\author{Ga\v{s}per Tka\v{c}ik}
\affiliation{Institute of Science and Technology, Am Campus 1, 3400 Klosterneuburg, Austria}

\begin{abstract}
    A key feature of many developmental systems is their ability to self-organize spatial patterns of functionally distinct cell fates. To ensure proper biological function, such patterns must be established reproducibly, by controlling and even harnessing intrinsic and extrinsic fluctuations. While the relevant molecular processes are increasingly well understood, we lack a principled framework to quantify the performance of such stochastic self-organizing systems. To that end, we introduce a new information-theoretic measure for self-organized fate specification during embryonic development. We show that the proposed measure assesses the total information content of fate patterns, and decomposes it into interpretable contributions corresponding to the positional and correlational information. By optimizing the proposed measure, our framework provides a normative theory for developmental circuits, which we demonstrate on lateral inhibition, cell type proportioning, and reaction-diffusion models of self-organization. This paves a way towards a classification of developmental systems based on a common information-theoretic language, thereby organizing the zoo of implicated chemical and mechanical signaling processes.
\end{abstract}

\maketitle

From the first cell division to the fully formed embryo, developmental systems exhibit a remarkable ability to generate functionally distinct cell types with precise positioning, timing and proportions. This process typically starts by establishing patterns of gene expression via cell-cell signaling, followed by the specification of individual cells into a cell fate. To achieve identical body plans across individuals, these patterns of cell fates must be reproducible~\cite{Wolpert1969}. In many classes of animals, including worms, insects and amphibia, the initial symmetry breaking to establish such patterns is driven by external input signals~\cite{Jukam2017}. For example, in the early fly embryo, morphogens produced by the fly mother break the initial symmetry and their precise establishment and readout are responsible for the developmental reproducibility~\cite{Gregor2007,Dubuis2013,Petkova2019}. In contrast, a broad variety of developmental systems break symmetry autonomously and self-organize patterns of gene expression. This includes the development of the early human embryo, where cells remain indistinguishable until the 8-cell stage, followed by self-organized polarity establishment~\cite{Wennekamp2013}. Furthermore, recently developed experimental \textit{in vitro} stem-cell patterning systems, such as two-dimensional assemblies~\cite{Warmflash2014}, intestinal organoids~\cite{Sato2009,Serra2019} and gastruloids~\cite{Beccari2018}, reveal a striking self-organization capability mimicking various stages of development. In all these examples, the system starts from an initially homogeneous, isotropic assembly of cells and -- in the absence of spatially structured external signals -- spontaneously breaks symmetry and self-organizes cell fate patterns. 

Self-organization in biological systems relies on the collective communication of constituent cells. This communication is implemented by different biophysical processes, including cell-cell signaling, morphogen dynamics, as well as cellular force generation and tissue mechanics~\cite{Schweisguth2019,Zinner2020a,Collinet2021,Kicheva2023}. To understand how these processes combine to control self-organization, mechanistic physical models, including reaction-diffusion, dynamical systems, and mechano-chemical models, have been developed~\cite{Green2015,Bailles2022a}. Nevertheless, the broad variety of mechanisms -- which often lead to qualitatively different final states -- has hampered the search for generally applicable principles and a common quantitative language for self-organization. For instance, while gastruloids self-organize by establishing positional patterning along an axis~\cite{Beccari2018}, intestinal organoids break symmetry by switching the fate of a single cell~\cite{Serra2019}, and the early human embryo initially patterns its inner cell mass into accurate cell type proportions without positional order~\cite{Saiz2016}. While precision in individual systems has been assessed by focusing on their idiosyncratic features, e.g., gene expression boundaries or cell type proportions, a generic, mechanism- and system-independent measure of reproducible self-organized cell fate patterning is currently lacking.

Self-organizing patterning processes generate, transmit, transform, and distribute information in space and time. Much as physics equips us with a formalism to describe how the flows of matter generate patterns, information theory provides a formal language to quantify statistical structures present in such patterns. In stochastic input-output systems driven by external inputs, performance can be defined in terms of the mutual information between input and output signals~\cite{Tostevin2009,Tkacik2016,Reinhardt2023}, which is maximized for optimally efficient information transmission channels~\cite{Tkacik2008}. This approach has previously been applied to formalize the notion of ``positional information'' in developmental systems~\cite{Dubuis2013,Tkacik2015,Tkacik2021a}, where inputs are the maternally provided morphogen gradients. However, for self-organizing patterns, input signals are either absent or not very expressive, and thus a more general approach is needed to quantify their information content.

Here, we address this challenge by proposing an information-theoretic measure of self-organization performance in embryonic development, which we introduce in Section~\ref{sec_framework}. Section~\ref{sec_dynsys} defines developmental processes in the language of stochastic dynamical systems. The main focus of this paper is the application of our information-theoretic approach to three paradigmatic stochastic models of self-organized patterning: lateral inhibition signaling (Section~\ref{sec_lis}), cell type proportioning and sorting (Section~\ref{sec_ex_prop}), and reaction-diffusion dynamics (Section~\ref{sec_ex_rd}). For all these systems we identify optimal parameter regimes where cell fates can emerge reproducibly in the presence of fluctuations, and where, furthermore, these fates can be locked into spatial orderings that correspond to precise body plans. 

\begin{figure}[h!]
	\includegraphics[width=0.48\textwidth]{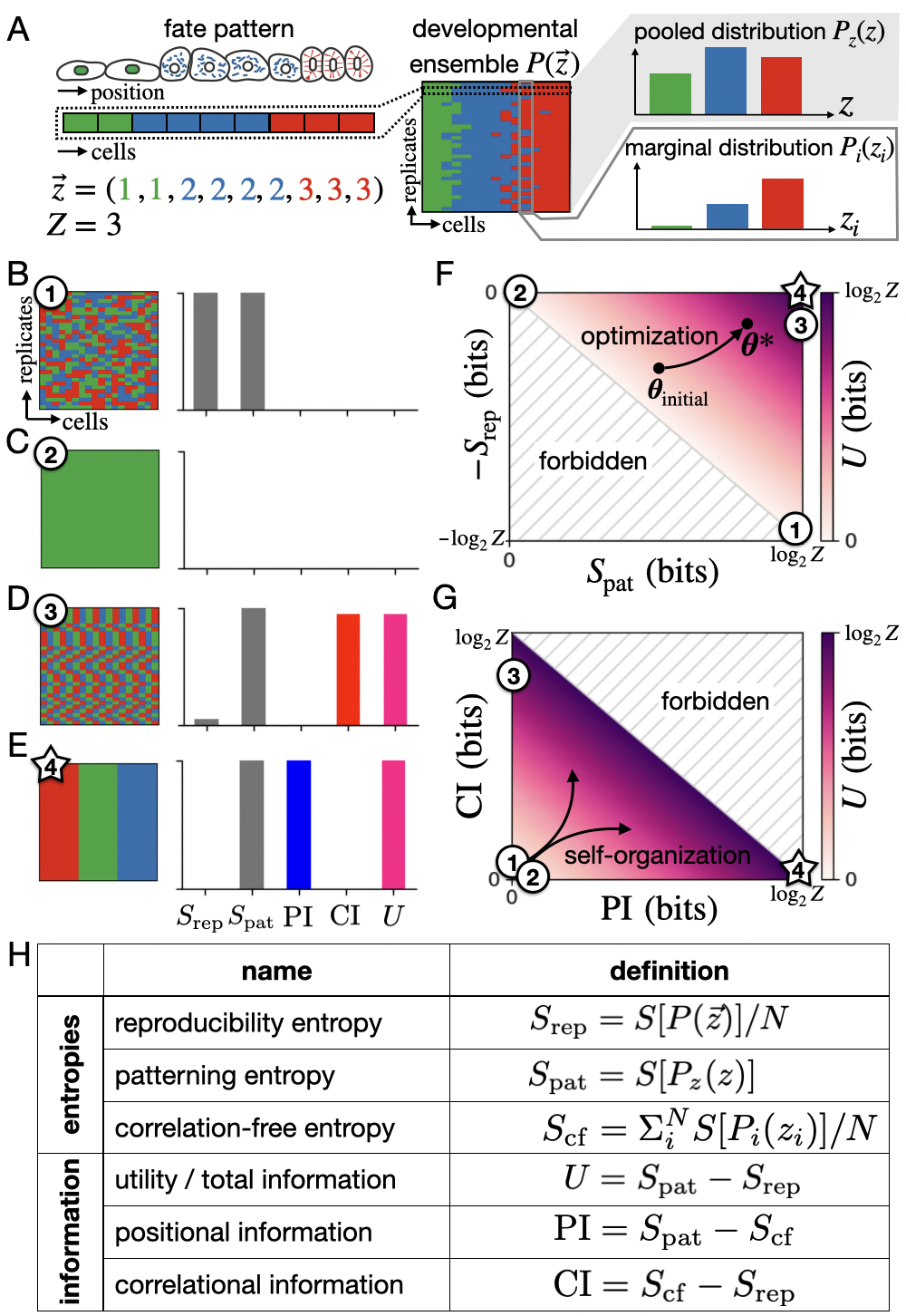}
	\centering
		\caption{\textbf{Entropy and information plane characterize self-organization outcomes.} 
            (\textit{A}) Schematic of the statistical approach to cell fate patterns, in which an assembly of cells of different cell fates is represented as a vector of discrete fates (left). Cells may have different shapes, sizes, and may be placed in complex, not necessarily one-dimensional, spatial arrangements. Sampling from the developmental ensemble $P(\vec{z})$ results in a list of replicates (middle). The ensemble is characterized by the distribution of fates $P_z(z)$ pooled across systems and positions; 
            and the marginal distribution $P_i(z_i)$ at each position (right).
            (\textit{B--E}) Examples of ensembles, their entropy values, and information content.
            (\textit{F}) Entropy plane spanned by the patterning and reproducibility entropies.
            (\textit{G}) Information plane spanned by the positional and correlational information contributions to the utility.
            (\textit{H}) Overview of the three key entropic quantities, and the three information quantities obtained by combinations of the entropies.
		 }
	\label{fig1}
\end{figure}

\section{Information-theoretic framework for self-organized cell fate patterns}
\label{sec_framework}

\subsection*{Utility function for self-organization}

Self-organization refers to phenomena where elementary constituents of a system interact with each other to create system-wide spatio-temporal ordering---in other words, a ``pattern.'' Self-organized patterning typically fulfills two criteria: (1) starting from an initially homogeneous state, the system generates patterns in absence of external (spatially structured) input, except for various sources of noise, such as random initial conditions and intrinsic stochasticity; (2) patterning occurs reproducibly, meaning that multiple replicates of the system self-organize into similar final patterns. This second criterion is fundamental to the biological function of development: to build a reliable body plan, patterning processes must achieve high levels of reproducibility of cell fate assignments across embryos. Note that this notion of reproducibility refers to the biological reproducibility of the system, rather than a measure of experimental or technical reproducibility.

Mathematically, criteria (1) and (2) can be subsumed by a single utility function, such that ``self-organized'' systems will be the ones that tend to optimize the utility; and the evaluation of this utility over patterns generated by some system can serve as a quantification of the self-organizing capability of the system. Specifically, we consider a very general class of developmental mechanisms that establish patterns of chemical and/or mechanical signals through interactions between cells (defined more precisely in Section~\ref{sec_dynsys}). These patterns are then interpreted by each single cell to specify a discrete cell fate. For a system (such as an embryo or an organoid), which at a particular developmental stage is composed of $N$ cells, we represent the fate pattern of each replicate as a vector
\begin{equation}
\label{eq_zvec}
\vec{z} = ( z_{1}, ..., z_{i}, ..., z_{N} ),
\end{equation}
where $z_{i} \in \{ 1,...,Z \}$ is the fate of cell $i$ chosen among $Z$ possible fates. Here, the index $i$ enumerates the cells, where the indices $i$ are tied to cell positions $\vec{x}=( \mathbf{x}_{1}, ..., \mathbf{x}_{i}, ..., \mathbf{x}_{N} )$, which are not necessarily one-dimensional. Since there is some freedom in how this indexing should be done, we here adopt a convention where we use global symmetries of the system (such as periodic boundaries or left-right symmetry) to align patterns where possible. 

An ensemble of fate patterns represents replicate outcomes of a developmental process, such as a collection of embryos (representative of a natural population) or of organoids, subject to naturalistic sources of noise and variability (Fig.~\ref{fig1}A). A typical patterning process will result in fate patterns that share similar features, but are not always identical. We can think of these replicates as samples from a joint probability distribution $P(\vec{z})$, which we refer to as the \textit{developmental ensemble}. To measure how reproducible these patterns are, we consider the entropy of the developmental ensemble, or \textit{reproducibility entropy}:
\begin{equation}
\label{eq_Srep}
S_\mathrm{rep} = \tfrac{1}{N} S \left[ P(\vec{z}) \right] = - \tfrac{1}{N} \textstyle\sum_{\vec{z}} P(\vec{z}) \log_2 P(\vec{z}).
\end{equation}
Reproducibility entropy has a minimum value of $S_\mathrm{rep}=0$ bits, corresponding to perfect reproducibility, which is realized by an ensemble of identical replicates; and a maximum value $\log_2 Z$, which is realized by a maximally irreproducible ensemble where all possible fate patterns have equal probability. Thus, a system with four possible fates can have at most 2 bits of reproducibility entropy.

Reproducibility alone is not sufficient to define self-organization: a system without any pattern can be perfectly reproducible. To measure the diversity of realized cell fates in the system, we first construct the pooled distribution $P_z(z) = \frac{1}{N} \sum_{i=1}^N \sum_{z_i=1}^Z P(\vec{z}) \delta(z_i,z)$ (Fig.~\ref{fig1}A), where $\delta(z_i,z)$ is the Kronecker delta. One can think of $P_z(z)$ as a distribution over cell fates in the entire developmental ensemble, i.e., as if one dissociated and pooled all cells together across positions and replicates. To quantify the patterning diversity, we define the \textit{patterning entropy}:
\begin{equation}
\label{eq_Spat}
S_\mathrm{pat} = S[P_z(z)] = -\textstyle\sum_{z=1}^Z P_z(z) \log_2 P_z(z).
\end{equation}
This entropy provides a scale for pattern diversity: if all cells have the same fate (no pattern), $S_\mathrm{pat}=0$ bits, while for equal proportions of all available fates, $S_\mathrm{pat}=\log_2 Z$ bits. In a system with four cell fates, the maximum patterning entropy is 2 bits.

Based on these definitions, we can formalize our two criteria for self-organization: a self-organizing system should seek to minimize $S_\mathrm{rep}$ while maximizing $S_\mathrm{pat}$. A utility function that is maximized by a self-organizing system can therefore be written as:
\begin{equation}
\label{eq_utility}
U = S_\mathrm{pat} - S_\mathrm{rep}.
\end{equation}
Clearly, specific biological systems may have been evolutionarily selected to produce very particular spatial patterns rather than just favoring any sufficiently diverse pattern. Similarly, we could require patterns to exhibit a certain degree of complexity, and could formulate alternatives to our utility to favor such outcomes (see Discussion).  

Such additional requirements are, however, unlikely to be generic. Furthermore, our proposed utility function would identify those more complicated outcomes as well, with additional biological or evolutionary considerations breaking the degeneracy between candidate reproducible patterns. Put conversely, systems whose utility is zero or vanishingly small cannot reasonably self-organize, no matter what non-trivial fate pattern is desired. Indeed, whatever specific pattern  may be optimal for the system at hand, the fundamental trade-off between being reproducible across replicates while creating diversity of cell types is general. Thus, in the absence of more specific constraints, Eq.~\eqref{eq_utility} provides a general and agnostic formulation of this trade-off.

\begin{figure*}
	\includegraphics[width=\textwidth]{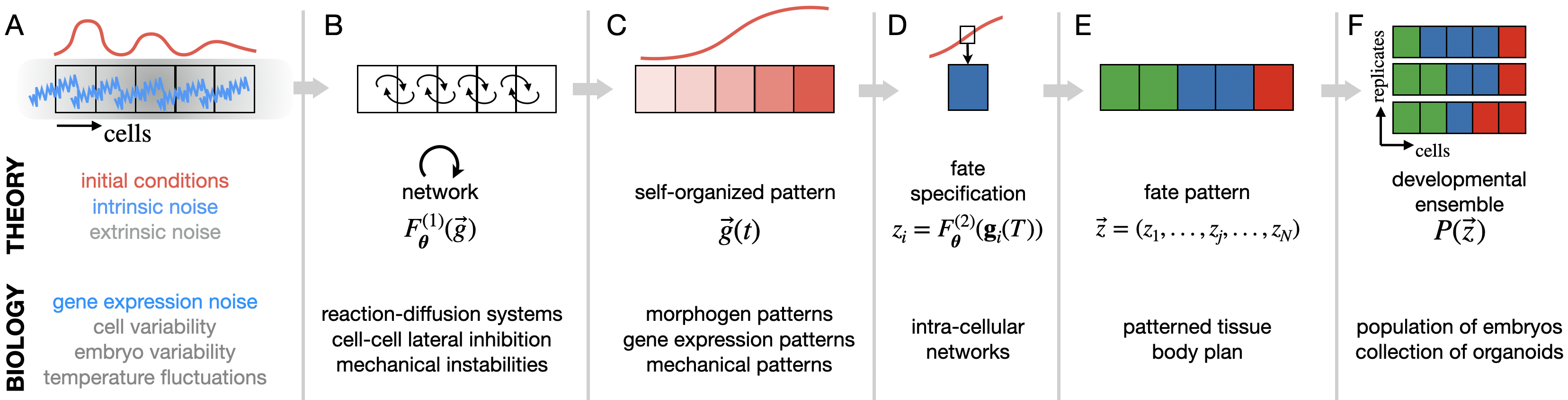}
	\centering
		\caption{\textbf{Cell fate patterning processes.} We describe cell fate patterning as a sequence of steps, shown as a schematic (top), with the corresponding description in our theoretical approach (middle), and possible biological implementations (bottom).
        (\textit{A}) The system is subject to various sources of stochasticity, including intrinsic noise and extrinsic noise across cells or replicates. The dynamics can also start with randomness in initial conditions. 
        (\textit{B}) The cells subsequently signal to each other through the signaling network determined by the dynamical systems specified by Eq.~\eqref{eq:dynsys}.
        (\textit{C}) This communication  establishes self-organized patterns $\vec{g}(t)$.
        (\textit{D}) Each  cell autonomously interprets the patterning concentrations at readout time $T$ to decide its fate $z_i$.
        (\textit{E}) Fate decisions of all cells yield the fate pattern of one replicate, $\vec{z}$.
        (\textit{F}) A large number of replicates constitutes the developmental ensemble $P(\vec{z})$.
		 }
	\label{fig2}
\end{figure*}

An important feature of our utility function is that it trades off the two entropies on equal terms, rather than using a trade-off parameter. This ensures that the utility has a lower bound at $U=0$ that is realized for any system generated by random, uncorrelated assignments of fates drawn from the pooled distribution $P_z(z)$ (Fig.~\ref{fig1}B). The fate pattern distribution corresponding to this scenario is the maximum entropy distribution given the observed pooled distribution,
\begin{equation}
\label{eq_maxent}
Q(\vec{z}) = \textstyle\prod_{i=1}^N P_z(z_i),
\end{equation}
for which $S_\mathrm{rep} = S_\mathrm{pat}$. This construction allows us to rewrite the utility (Eq.~\eqref{eq_utility}) as a Kullback-Leibler (KL) divergence between the observed distribution $P$ and the maximum entropy distribution $Q$ (see Supplementary Information):
\begin{equation}
\label{eq_DKL}
U = \tfrac{1}{N} D_\mathrm{KL} \left[ P(\vec{z}) || Q(\vec{z}) \right] = \tfrac{1}{N} \textstyle\sum_{\vec{z}} P(\vec{z}) \log_2 \left( \frac{P(\vec{z})}{Q(\vec{z})} \right).
\end{equation}
Evidently, the state of no patterning has zero utility, since in that case $P=Q$ by definition, and  the divergence vanishes (Fig.~\ref{fig1}C).

Developmental ensembles produced by patterning systems can be visualized in the \textit{entropy plane}, the two-dimensional space spanned by the two entropies, $S_{\rm rep}$ and $S_{\rm pat}$. This plane is bisected by a diagonal defined by the limit of maximal irreproducibility (Fig.~\ref{fig1}F). Due to the equal trade-off of the two entropies, this limit corresponds to minimal utility and all lines of constant utility are parallel to the diagonal; the optimum at maximal utility of $U=\log_2 Z$ bits is in the top right corner. This optimum corresponds to a system with high patterning diversity and perfect reproducibility, such as a perfect ``flag''-pattern (Fig.~\ref{fig1}E). In summary, the utility scores the outputs of any possible patterning mechanism onto a unique quantitative scale, without reference to the underlying mechanisms. Conversely, optimal parameters of a patterning process can be identified by using the utility as an optimization function. 

\subsection*{Decomposition into positional and correlational information}

Reproducibility is achieved by tightening the joint probability distribution $P(\vec{z})$ in the high-dimensional space of cell fate assignments, raising the question of how to conceptualize the information contained in such a distribution. The width of the probability distribution of cell fates at each position $i$ -- the marginal distribution $P_i(z_i)$ -- determines how much local information is contained in the pattern, i.e. how much information the position $i$ carries about the fate $z_i$ and vice versa (Fig.~\ref{fig1}A). However, reproducibility entropy can also be reduced through correlations from position to position. What is the additional information contained in such correlations?

We can formalize this question by decomposing the utility into two interpretable quantities. First, tightening the marginal distributions while maximizing the patterning entropy corresponds to maximizing the \textit{positional information} (PI) of the pattern~\cite{Dubuis2013,Tkacik2015,Tkacik2021a}. Specifically, since indices are tied to positions in space, the PI of an ensemble of fate patterns is given by the mutual information of cell fate $z$ and cell index $i$: 
\begin{align}
\label{eq_PI}
\mathrm{PI} &= \textstyle\sum_{i=1}^N \textstyle\sum_{z=1}^{Z} P(z,i) \log_2 \left( \frac{P(z,i)}{P_z(z)P_\mathrm{index}(i)} \right)
\end{align}
where $P(z,i)$ is the joint distribution of cell fates and indices, averaged over the developmental ensemble. Using Bayes' rule $P(z,i)=P_i(z_i)P_\mathrm{index}(i)$ and the fact that the indices are by definition uniformly distributed, $P_\mathrm{index}(i)=1/N$, this simplifies to (Supplementary Information):
\begin{align}
\label{eq_PI_entropies}
\mathrm{PI} &= S_\mathrm{pat} - S_\mathrm{cf},
\end{align}
where we defined the \textit{correlation-free entropy},
\begin{equation}
\label{eq_Scf}
S_\mathrm{cf} = \tfrac{1}{N} \textstyle\sum_{i=1}^N S[P_i(z_i)],
\end{equation}
which is the entropy of a joint distribution constructed from independent marginals, i.e., $P(\vec{z}) =\prod_{i=1}^N P_i(z_i)$, corresponding to a system with no spatial correlations. We can now compute the reduction in entropy due to the presence of spatial correlations, which we term \textit{correlational information} (CI):
\begin{equation}
\label{eq_dS}
\mathrm{CI} = S_\mathrm{cf} - S_\mathrm{rep}.
\end{equation}
This information can be rewritten as a KL divergence between the developmental ensemble and the correlation-free distribution, $\mathrm{CI} = \tfrac{1}{N} D_\mathrm{KL}\left[ P(\vec{z}) || \prod_{i=1}^N P_i(z_i) \right]$. This measure of correlation in a joint probability distribution has been previously defined in the information-theoretic literature, and is often referred to as the ``multi-information'' of the distribution~\cite{Watanabe1960,Studeny1998}.

Combining~Eq.~\eqref{eq_PI} and~Eq.~\eqref{eq_dS}, we find that
\begin{equation}
\label{eq_DKL_PIdS}
U = \mathrm{PI} + \mathrm{CI}.
\end{equation}
The utility can therefore be understood as the sum of two non-negative contributions, the local (positional) and the non-local (correlational) information, that together provide a quantification of the total information content of an ensemble of patterns.

To gain intuition for this decomposition, a helpful geometric construction is to consider the \textit{information plane} spanned by the positional and correlational information (Fig.~\ref{fig1}G). Unlike the entropy plane, the information plane has a unique location for the minimum of the utility, where both information terms vanish; and a broad range of possible combinations of the two terms that result in similar utility values. Both terms vanish for uniform or maximally disordered ensembles (Fig.~\ref{fig1}B,C). Systems with strong correlation of cell fate to cell position, and therefore low-entropy marginal distributions, have high PI (Fig.~\ref{fig1}E). However, patterns with low PI may still contain significant structure. For instance, a perfect alternating pattern with random shifts has zero PI, but high CI (Fig.~\ref{fig1}D). Self-organization can thus proceed by (1) setting up correlations of gene expression with position and/or by (2) setting up correlations across positions. Note that the state of maximum utility necessarily corresponds to maximum PI, as the only way to \emph{globally} maximize the utility is if all replicates are identical, implying $U = \mathrm{PI} = \log_2 Z$ and $\mathrm{CI}=0$, by construction. On the one hand, this result is a mathematical necessity by virtue of our careful definitions; on the other, however, its biological significance is deeply non-trivial: if a biological system can achieve values of utility close to their maximal bound, the \emph{only} way to do so is to maximize PI, i.e., to generate a reproducible body plan. 

Taken together, our framework identifies three entropies to quantify patterning, which combine to give three quantities to estimate the total, the positional, and the correlational information. All six quantities are summarized in Fig.~\ref{fig1}H. 

\begin{figure*}
\centering
\includegraphics[width=0.7\textwidth]{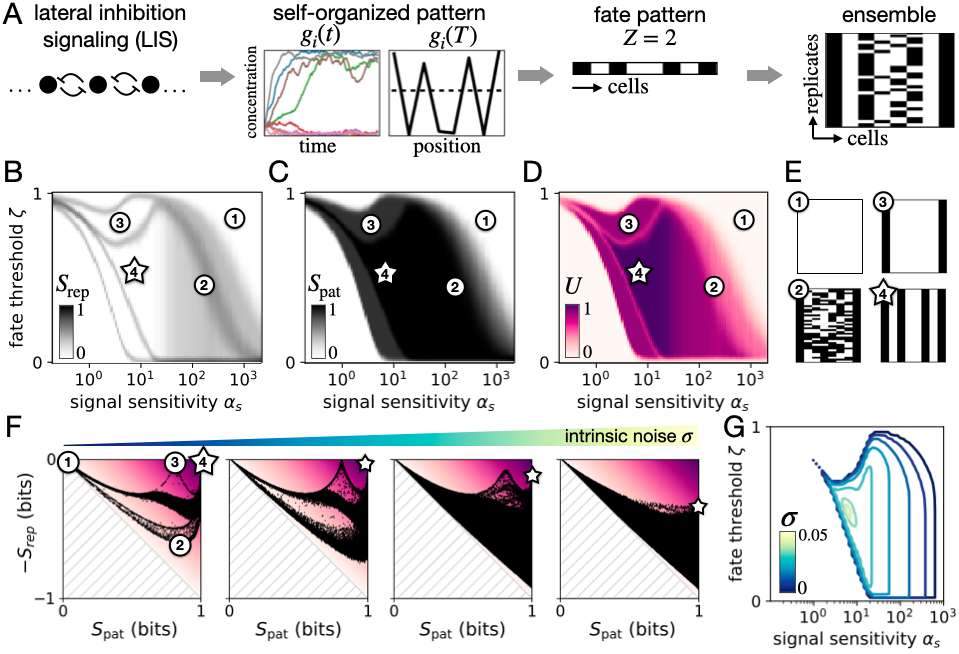}
\caption{\textbf{Optimal patterning in a minimal stochastic lateral inhibition system.} 
        (\textit{A}) The production of chemical $g$ in each cell is subject to inhibition by neighbouring cells with a sensitivity parameter $\alpha_s$. We simulate $N=8$ cells in 1D with nearest neighbor interactions and closed boundary conditions. As time unfolds, symmetry is broken with some cells having a high and some cells a low concentration of $g$. These concentrations are thresholded with threshold $\zeta$ at readout time $T$ into $Z=2$ cell fates (depicted with black and white in the developmental ensemble).
        (\textit{B,C,D}) Reproducibility entropy, patterning entropy, and utility, respectively, as a function of $\alpha_s$ and $\zeta$. Numbers (1)-(4) denote example ensembles shown in panel E.
        (\textit{E}) Depictions of four developmental ensembles: no patterning (1), maximally irreproducible ensemble (2), boundary cells only (3), and reproducible alternating patterning (4).
        (\textit{F}) Visualization of patterning outcomes in entropy planes for four increasing intrinsic noise levels ($\sigma=\{0.001,0.01,0.05,0.1\}$). Each of the $10^5$ dots corresponds to a developmental ensemble defined by a random draw of its parameters $\boldsymbol{\theta}=\{ \alpha_s,\zeta \}$.
        (\textit{G}) High utility regions in parameter space, defined as $U(\boldsymbol{\theta})>0.99 U(\boldsymbol{\theta}^*)$, for various noise levels $\sigma$ (color-coded).
		 }
	\label{fig_LIS}
\end{figure*}

\section{Self-organized patterning as a stochastic dynamical system}
\label{sec_dynsys}

To illustrate how our framework can be applied, we consider a general dynamical process that governs a stage of development which, after a finite time $T$, gives rise to a spatial pattern of cell fates. Our aim is to quantify the performance of self-organized patterning  at this readout time. For times $t\in [0,T]$, we consider a very generic implementation of a chemical reaction network, which could include gene regulatory and cell signaling dynamics, cell-to-cell coupling, as well as cell divisions and apoptosis. At the readout time, we take the system to be composed of $N$ discrete cells. Here, we fix the cell number $N$ throughout the developmental stage for convenience, but this simplification can be relaxed. The state of each cell $i$ at time $t$ is described by the chemical concentration vector $\mathbf{g}_{i}(t)$. We assemble the state of each replicate into a concentration vector 
\begin{equation}
\vec{g}(t) = ( \mathbf{g}_{1}(t), ..., \mathbf{g}_{i}(t), ..., \mathbf{g}_{N}(t) ).
\end{equation}
The regulatory dynamics of each cell are described by a stochastic dynamical system (Fig.~\ref{fig2}B)
\begin{equation}
\label{eq:dynsys}
\frac{\partial \mathbf{g}_{i}}{\partial t} = F_{\boldsymbol{\theta}}^{(1)}(\vec{g}) + \sigma(\mathbf{g}_{i}) \xi(t).
\end{equation}
where $\xi(t)$ is a multivariate zero-mean unit-covariance Gaussian white noise process. We allow for a state-dependent magnitude $\sigma(\mathbf{g}_{i})$ to model, for example, multiplicative gene expression noise. The dynamical system $F_{\boldsymbol{\theta}}^{(1)}(\vec{g})$ is a general non-linear function that describes spatial coupling, chemical reactions, and cell-cell interactions, and is determined by a set of parameters $\boldsymbol{\theta}$.

To investigate the performance of self-organizing systems, we treat noise as an integral part of the problem, since it imposes constraints and trade-offs on signaling mechanisms which need to be navigated to achieve final states of high utility. We therefore focus on self-organization of intrinsically stochastic systems and consider the following sources of noise (Fig.~\ref{fig2}A): (1) \textit{noise in the initial conditions}, i.e. in $\vec{g}(t=0)$, (2) \textit{intrinsic noise} $\xi(t)$ with state-dependent magnitude $\sigma(\mathbf{g}_{i})$ due to thermal and small number fluctuations in mechano-chemical processes (Eq.~\eqref{eq:dynsys}), and (3) \textit{extrinsic noise}, which subsumes cell-to-cell and replicate-to-replicate (i.e. embryo-to-embryo) variability in the parameters, such as size or temperature fluctuations across embryos. This means that parameters are drawn from a distribution $P_{\boldsymbol{\theta}}({\boldsymbol{\theta}};{\boldsymbol{\bar{\theta}}},\sigma_{\boldsymbol{\theta}})$, which we will assume to be Gaussian in our examples below. Here, the magnitude of the variability $\sigma_{\boldsymbol{\theta}}$ is held fixed as a constraint and we seek to optimize ${\boldsymbol{\bar{\theta}}}$.

The concentrations of the signaling molecules eventually trigger the specification of each cell into a discrete cell fate. Thus,  while the continuous concentration variables $\mathbf{g}_i$ can be spatially coupled, the commitment to a discrete cell fate typically includes cell-autonomous decision points where an individual cell $i$, based on the local concentrations $\mathbf{g}_i$, locks into a particular ``memory'' state (the cell fate) (Fig.~\ref{fig2}C, D). The cell-autonomy of this step is motivated by the fact that cell fate decisions are typically implemented at the level of the genome, i.e. within each cell nucleus. To this end, we include a fate decision step that maps  continuous concentration levels in each cell $\mathbf{g}_i$ to a discrete fate $z_i$. This step is defined by the fate decision function $z_i=F^{(2)}_{\boldsymbol{\theta}}(\mathbf{g}_i(t))$. This formulation encompasses multiple ways of thinking about cell fate decisions. For example, $F^{(2)}$ could be implemented as a set of thresholds acting on a single concentration at the final time point $g_i(T)$, corresponding to the classical Wolpertian ``French-Flag'' picture~\cite{Wolpert1968}. More generally, the decision function could combine multiple input concentrations non-linearly to select a cell fate~\cite{Zagorski2017}.
Alternatively, the decision step could be implemented via a dynamical system with discrete attractors, which can be seen a mathematical realization of the Waddington landscape proposed in previous work~\cite{Corson2012,Rand2021}. Irrespective of the details, in all these cases each cell will autonomously convert continuous inputs into a discrete cell fate. Note that the decision function is also parameterized by the parameter vector $\boldsymbol{\theta}$, and is therefore subject to optimization of the utility function. This final fate decision step yields the discrete fate pattern in each replicate, $\vec{z}$ (Fig.~\ref{fig2}E,~Eq.~\eqref{eq_zvec}). Based on an ensemble of such fate patterns (Fig.~\ref{fig2}F), we can then evaluate the utility with Eqs.~(\ref{eq_Srep}--\ref{eq_utility}).

\section{Example 1: Lateral Inhibition Signaling}
\label{sec_lis}

In our first example, we explore the optimal performance of lateral inhibition signaling (LIS). Biologically, LIS is realized by the Delta-Notch pathway~\cite{Sprinzak2010}, which plays a key role in a broad range of self-organized developmental systems~\cite{Cohen2010,Hadjivasiliou2016,Corson2017b,Serra2019,Gozlan2023}. In a minimal model of this pathway, each cell $i$ produces a chemical $g_i$, which inhibits the production of the same chemical in neighbouring cells~\cite{Collier1996}. We start with a simple stochastic production-degradation dynamics:
\begin{equation}
\label{eq_LIS}
\frac{\mathrm{d}g_i}{\mathrm{d}t} = \frac{1}{1+\exp[-f(g_i,s_i)]} - g_i + \sigma \xi(t),
\end{equation}
where $f(g_i,s_i)$ is a generic regulatory function that controls the production of $g$ in response to multiple inputs: $g$ itself as well as the signal received from neighbouring cells, $s_i$. Different choices for $f$ correspond to a range of previously studied models (Supplementary Information). The signal sent by neighbouring cells is $s_i = \sum_{j\neq i} c_{ij} g_j$, where $c_{ij}$ is the connectivity of the cells, here taken according to a 1D ordering and closed boundaries (i.e. $c_{ij}=1$ for $j=i\pm1$, and $c_{ij}=0$ otherwise). 

We first consider a single regulatory input to $f$, i.e., the signal $s_i$ from neighbouring cells. The simplest choice corresponds to $f(g_i,s_i) = -\alpha_s s_i$, which, for sensitivity $\alpha_s > 0$, will result in the desired repression of $g_i$ by neighboring cells. In the presence of stochasticity, here implemented through the intrinsic noise $\xi(t)$ with magnitude $\sigma$, lateral inhibition leads to symmetry breaking into cells with high and low concentrations of $g$. The fate decision function in this example is simply a deterministic binary threshold $z_i = H(g_i(T)-\zeta)$ where $H$ is a Heaviside step function and $\zeta$ is the threshold parameter. Thus, intracellular concentrations are thresholded into patterns of two cell fates, with some variability across replicates (Fig.~\ref{fig_LIS}A). This example can be seen as an instantiation of the general scheme shown in Fig.~\ref{fig2}.

\begin{figure}
	\includegraphics[width=0.48\textwidth]{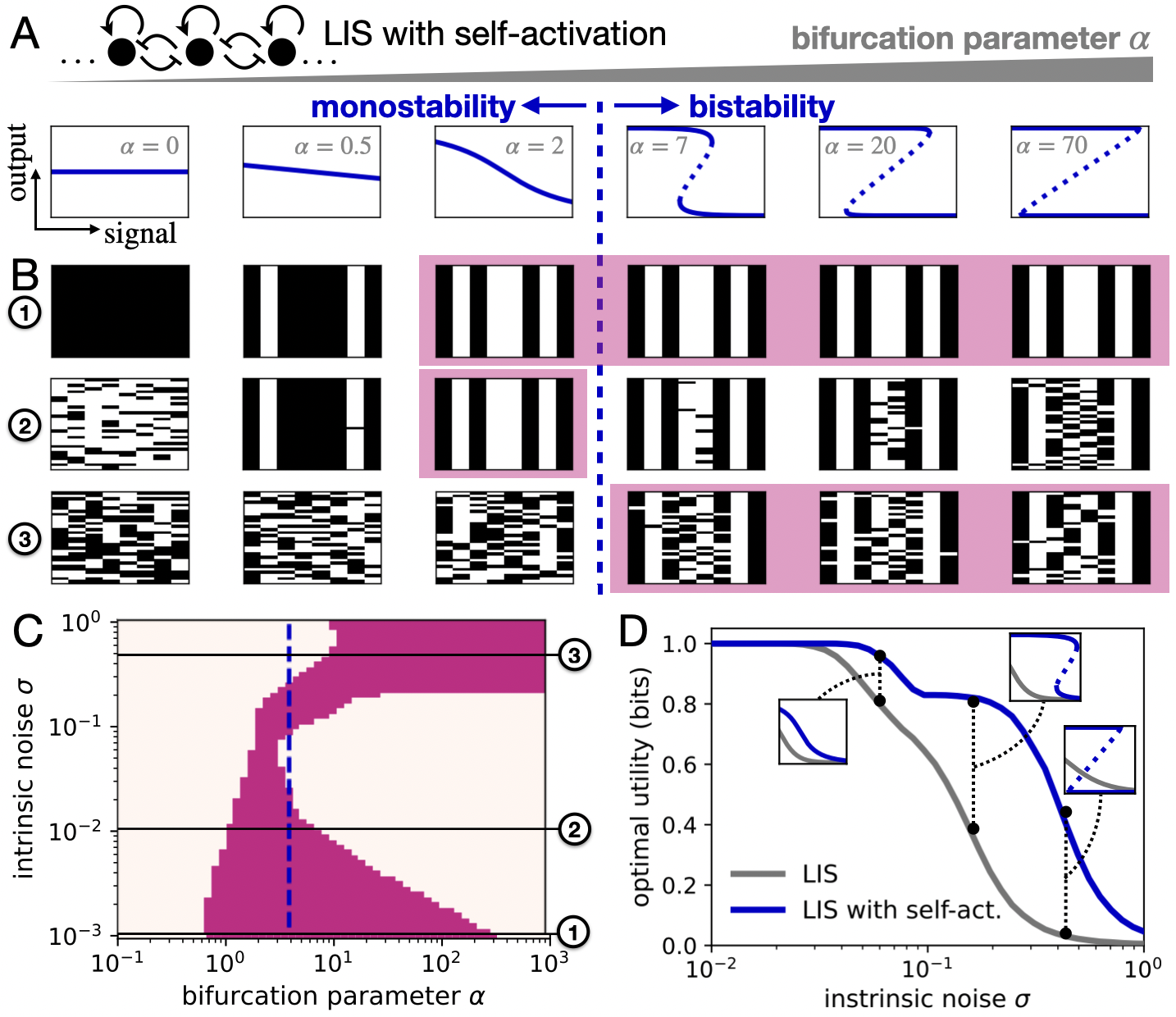}
	\centering
		\caption{\textbf{Optimal patterning in a stochastic lateral inhibition system with self-regulation.}
        (\textit{A}) Nullclines of Eq.~(\ref{eq_LIS}) as a function of the bifurcation parameter $\alpha$. Solid and dotted lines correspond to stable and unstable fixed points, respectively. Dashed line between panels indicates the onset of bistability at critical $\alpha_\mathrm{c}=4$.
        (\textit{B}) Developmental ensembles for different bifurcation parameter values and nullclines as shown in A above.  Rows (1), (2), (3)  correspond to three intrinsic noise levels $\sigma=\{ 0.001,0.01,0.4 \}$. Purple frames indicate the optimal parameter ranges at each noise level, as in C.
        (\textit{C}) Optimal parameter range for $\alpha$ as a function of noise, defined as those values of $\alpha$ that have at least 90\% of the maximum utility at each noise level.
        (\textit{D}) Utility of optimal LIS models without (gray) and with (blue) self-regulation, as a function of noise. For each noise level, all parameters of both LIS models were optimized independently. Insets show optimal response curves of both models at three example noise levels.
        All panels are plotted with optimized thresholds $\zeta^*$.
		 }
	\label{fig_LIS_bistable}
\end{figure}

Our minimal model for LIS only has two parameters, $\boldsymbol{\theta}=\{ \alpha_s,\zeta \}$, yet already generates a complex space of patterning outcomes with different apparent levels of reproducibility. Entropy calculations confirm this observation and identify three patterns with nearly perfect reproducibility ($S_\mathrm{rep} \approx 0$ bits), corresponding to no patterning, specification of boundary cells only, and alternating patterns (Fig.~\ref{fig_LIS}B,E). Positive values for patterning entropy rule out trivial outcomes and quantitatively distinguish between high-diversity patterns, such as the alternating solution with $S_\mathrm{pat} = 1$ bit, and lower-diversity patterns, such as the boundary cell specification with $S_\mathrm{pat} \approx 0.81$ bit (Fig.~\ref{fig_LIS}C,E). Combining both entropy terms into the utility function finally identifies the optimal parameters where reproducible, high-diversity patterns emerge at the maximum utility value of $U=1$ bit (star in Fig.~\ref{fig_LIS}D). 

In this example, the parameter space is low-dimensional and thus easily visualized. For complex networks with high-dimensional parameter spaces, direct visualization and parameter scans will no longer be feasible, but (stochastic) optimization of the utility can nevertheless be performed to identify optimal parameters, $\boldsymbol{\theta^*} = {\mathrm{argmax}}_{\boldsymbol{\theta}} \ U(\boldsymbol{\theta})$. An alternative approach is to visualize patterning outcomes in the two-dimensional entropy plane, which is possible regardless of the dimensionality of the model's parameter space. To demonstrate this, we randomly draw parameters $\boldsymbol{\theta}$ for our LIS model from a broad domain of values, and visualize the entropies corresponding to each draw (Fig.~\ref{fig_LIS}F). We recover the utility-maximizing alternating patterns in the top right corner of the entropy plane, as expected. The point cloud shape in the entropy plane is determined by the underlying mechanism, which constrains the possible patterning outcomes, as well as the discreteness of the system; it depends on $N,Z$ and the boundary conditions (Supplementary Information).

We next ask whether our framework can identify non-trivial utility optima, and how these optima vary as a function of noise in the system, which we treat as a parametric constraint. Indeed, as the noise is increased, the optimal region in parameter space shrinks (Fig.~\ref{fig_LIS}G). Specifically, there is a minimum sensitivity $\alpha_s^\mathrm{min}$ below which cells are insensitive to the received signal and no patterning occurs. While at vanishing noise the sensitivity can grow arbitrarily large with no detrimental effect on patterning, as noise is increased, a clear upper bound $\alpha_s^\mathrm{max}$ emerges. Above this bound, cells become too sensitive to the noise in the system, causing the utility to drop precipitously. Even with the values of parameters optimized independently for every choice of noise, the absolute value of the utility drops significantly as noise is increased (Fig.~\ref{fig_LIS_bistable}D, gray line). 

Optimization of our model invariably ensures maximal patterning entropy through optimal placement of the decision threshold. Thus, in this example, utility maximization reduces to maximization of reproducibility, subsuming the concept of optimizing robustness to noise~\cite{Barkai1997}. To investigate whether additional regulatory complexity can enhance the patterning performance at large noise levels, we extend our model by adding a self-activation term to $f$ and write $f(g_i,s_i) = -\alpha_s s_i + \alpha_g g_i$ with $\alpha_g>0$. Self-activation leads to a cell-intrinsic bistability, such that regulatory input from neighboring cells biases the focal cell into one of the two effective potential minima (attractors), each corresponding to a possible cell fate~\cite{Corson2017b}.

To gain intuition, we first set $\alpha_s = \alpha_g = \alpha$, which captures the essential phenomenology (Supplementary Information). In this case, $\alpha$ is a bifurcation parameter with a critical value $\alpha_\mathrm{c}=4$: intrinsic cell behavior switches from monostable for $\alpha<\alpha_\mathrm{c}$ to bistable for $\alpha>\alpha_\mathrm{c}$ (Fig.~\ref{fig_LIS_bistable}A). For small noise, utility is close to maximal across a broad range of $\alpha$ values: the minimum of this range at very small $\alpha$ corresponds to regulation functions that are too ``shallow'' to permit patterning; the maximum corresponds to excessive sensitivity to noisy signals that force the cells to stochastically transition to the ``wrong'' attractor~(Fig.~\ref{fig_LIS_bistable}A,B).
Interestingly, at intermediate noise ($10^{-2} \lesssim \sigma \lesssim 10^{-1}$), the utility optimum becomes increasingly narrow and peaks in the monostable regime just short of bifurcation (Fig.~\ref{fig_LIS_bistable}C); this may be due to efficient noise averaging afforded by the lengthening correlation time close to criticality~\cite{Tkacik2012a}. At larger noise, fluctuations destroy pattern reproducibility in the monostable regime due to the graded response of the cells to noisy inputs. Instead, the bistable regime becomes optimal, as the attractors protect the system from random fluctuations~\cite{Tkacik2010}. These results reveal a surprising non-monotonic dependence of optimal  parameters -- and even of qualitative optimal network behavior -- on the noise level, suggesting that bistability only enhances patterning reproducibility in the large noise regime. When noise is low, hysteresis causes fate decision defects, and graded responses are preferable to bistability. Finally, we demonstrate that optimizing the utility across the entire parameter space $\boldsymbol{\theta}=\{\alpha_s, \alpha_g, \zeta\}$ indeed substantially expands the range of noise amplitudes where high reproducibility patterns are attainable (Fig.~\ref{fig_LIS_bistable}D).

\begin{figure}[tb]
	\includegraphics[width=0.48\textwidth]{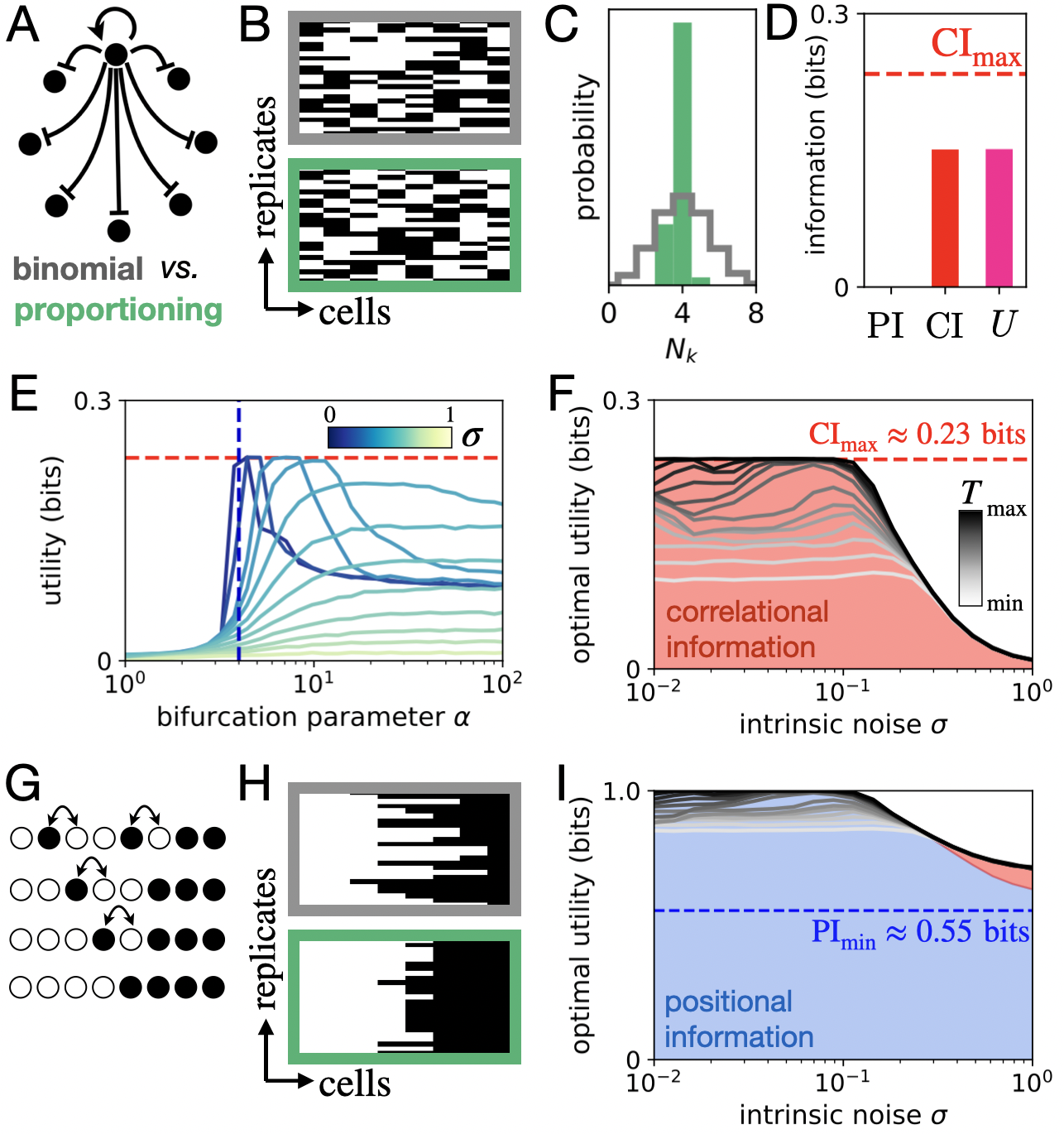}
	\centering
		\caption{\textbf{Optimization of cell-type proportioning and sorting.}
        (\textit{A}) Schematic of an all-to-all inhibition network with $N=8$ cells, including self-activation in each cell (interactions shown for a single focal cell). Eq.~\eqref{eq_LIS} with $f(g_i,s_i)=\alpha(g_i-s_i)$ is used.
        (\textit{B}) Developmental ensembles based on independent (binomial) cell fate decisions with equal probability $p=0.5$ for both fates (top, gray); or on proportioning mechanism of panel A with $\alpha=3, \sigma=0.15$ (bottom, green).
        (\textit{C}) Proportioning distribution $P(N_k)$ of the number of black cells $N_k$ per replicate, for the binomial ensemble (grey line) and the proportioning mechanism (green).
        (\textit{D}) PI, CI and $U$ for the proportioning mechanism. Red dashed line indicates the maximum possible CI for a proportioning process, $\mathrm{CI}_\mathrm{max}=\log_2 Z - \tfrac{1}{N}\log_2[N!/(N/Z)^Z]$ (assuming $N$ is divisible by $Z$, Supplementary Information).
        (\textit{E}) Utility as a function of the bifurcation parameter $\alpha$ for varying intrinsic noise level $\sigma$ (colorbar). Blue dashed line indicates  $\alpha_\mathrm{c}$; red dashed line indicates $\mathrm{CI}_\mathrm{max}$.
        (\textit{F}) Utility as a function of $\sigma$ with optimally chosen $\alpha$ at each noise level. Different greyscale curves are computed for different  values of readout time $T$ (colorbar).
        (\textit{G}) Schematic of the cell sorting process: cells in each replicate swap places until all black cells are on the right.
        (\textit{H}) Developmental ensembles obtained by noise-free sorting of the ensembles in B.
        (\textit{I}) Positional information as a function of intrinsic noise in the initial proportioning process, obtained after noise-free sorting of the ensembles in F. Blue dashed line: lower bound on PI obtained by sorting a binomial ensemble (Supplementary Information). Optimised thresholds $\zeta^*$ are used throughout.
		 }
	\label{fig_prop}
\end{figure}

\section{Example 2: Cell-type proportioning and sorting}
\label{sec_ex_prop}

In the lateral inhibition example, local cell-cell interactions established cell fate patterns with high positional order and thus high utility. Yet high utility can also arise when positional order is absent. This could happen, for instance, when a developmental system generates precisely controlled proportions of various cell types whose locations are, however, random across replicates. Known mechanisms that lead to such ``cell-type proportioning'' include lineage dynamics~\cite{Huelsz-Prince2022}, competition for a limited resource~\cite{Kitadate2019}, or long-ranged inhibitory interactions~\cite{Raina2021,Stanoev2021,Raju2023}. In some cases, proportioning is followed by ``cell-type sorting'', where different cell types segregate in space to form spatial patterns. Here we would like to understand pattern formation based on proportioning and sorting through the lens of our information-theoretic measures.

A simple implementation of the proportioning process employs long-range inhibition  to which cells respond with a cell-intrinsic bistability. Mathematically, a convenient model is \eqref{eq_LIS} introduced in the previous section, but with an all-to-all  connectivity ($c_{ij}=1$ for all $j\neq i$), as schematized in Fig.~\ref{fig_prop}A. This leads to self-organized proportioning into two equally represented cell types with no positional order (Fig.~\ref{fig_prop}B). Importantly, the distribution of cell fates can be substantially tightened around the ideal 1:1 split via the long-range inhibition, compared to the binomial expectation when cells do not signal to each other and decide their fates independently (Fig.~\ref{fig_prop}C). 
It is precisely this divergence between the developmental ensemble of interacting cells that are making correlated decisions, and the independent cells scenario, that is responsible for the substantial amount of correlational information (CI) and high utility even when positional information (PI) remains precisely at zero (Fig.~\ref{fig_prop}D).

For an idealized 1:1 position-independent proportioning  we can compute the theoretically maximal value of CI (Supplementary Information). We find that our minimal model can achieve this bound at sufficiently low noise if its bifurcation parameter $\alpha$ is optimized (Fig.~\ref{fig_prop}E). Interestingly, for a range of very low noise levels, the optimal $\alpha$ is close to the bifurcation point, $\alpha^* \sim \alpha_c$; as noise increases, it moves further into the bistable regime while still maintaining CI at the theoretical maximum (Fig.~\ref{fig_prop}F); until, for sufficiently large noise, optimization can no longer compensate and CI drops rapidly towards zero. Thus, while the shift into the bistable regime with increasing noise is similar to that observed for LIS (Fig.~\ref{fig_LIS_bistable}C), the global inhibition network in the proportioning process does not perform patterning at sub-critical values of $\alpha$. Future work will investigate how the underlying features of the super- and sub-critical dynamical systems relate to noise level~\cite{Tkacik2012a}.

Further analysis of the results in Fig.~\ref{fig_prop}F highlights the importance of timescales in our optimization framework. The baseline results assumed $T\rightarrow \infty$ to ensure that the utility function scores steady-state developmental ensembles. What happens, however, if the readout time $T$ is treated as a constraint that is progressively shortened? This is biologically relevant, since developmental steps usually follow a precise temporal schedule; scenarios where the readout happens before the system has fully relaxed have previously been suggested~\cite{Bergmann2007}. Indeed, as $T$ is shortened, the degeneracy over noise levels is lifted: an optimal noise level emerges at which the utility can be globally maximized with an appropriate choice of $\alpha$ (Fig.~\ref{fig_prop}F). This happens because favored values $\alpha \sim \alpha_c$ give rise to critical or bistable potential landscapes where dynamics drastically slows down or possibly gets stuck in a wrong attractor, such that a judicious addition of noise will help the system reach high-utility steady states even within a limited $T$.

After establishing two cell types in a proportioning process, a common developmental step is cell sorting into distinct spatial domains. This establishes positional order and thus generates PI, which we explore in a minimal model of proportioning followed by sorting that adds no further sources of stochasticity (Fig.~\ref{fig_prop}G,H). Sorting can establish PI even in absence of initial CI (by sorting draws from a binomial ensemble), thereby providing a lower bound to the generated PI (Fig.~\ref{fig_prop}I, Supplementary Information). Substantially higher PI (and utility) values can be reached, however, if sorting acts on an ensemble proportioned precisely with an optimal choice of $\alpha$. In this case, we find that PI closely mirrors the behavior of CI as a function of noise, both in steady-state as well as under limited $T$ scenarios (Fig.~\ref{fig_prop}F,I). Taken together, our framework suggests how evolution could co-opt low levels of biological noise and optimize a common, simple signaling system to improve patterning performance via interaction-driven cell proportioning and sorting -- namely, by generating correlational order and efficiently transforming it into positional information.

\begin{figure*}
	\includegraphics[width=0.9\textwidth]{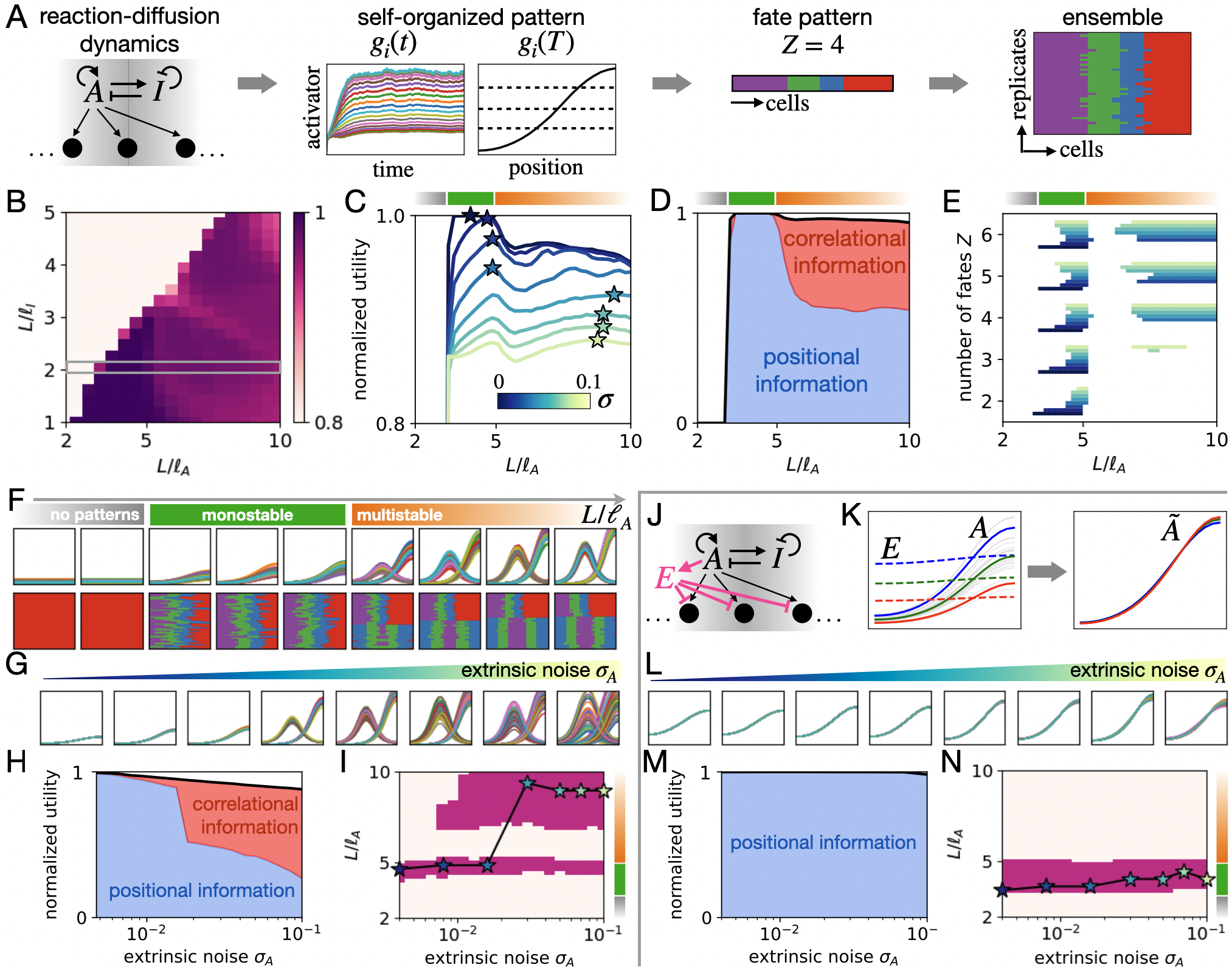}
	\centering
		\caption{\textbf{Optimal patterning in a stochastic activator-inhibitor system.} 
        (\textit{A}) Schematic of the cell fate patterning process in the activator-inhibitor system. Fate patterns are generated by thresholding the activator profile using a set of thresholds $\{\zeta_1,...,\zeta_{Z-1}\}$ for $Z$ fates.
        (\textit{B}) Utility as a function of the activator and inhibitor diffusive length scales ($\sigma_A=0$).
        (\textit{C}) Normalized utility as a function of $L/\ell_A$ with $L/\ell_I=2$, corresponding to the grey rectangle marked in B, for increasing extrinsic noise magnitude $\sigma_A$ using fixed $Z=4$. Shaded bars at the top correspond to the stability regimes indicated in F. Stars denote maximal utility solutions at each noise level.
        (\textit{D}) PI and CI contributions to the utility as a function of $L/\ell_A$ ($\sigma_A=0$).
        (\textit{E}) Top 1\% utility parameter regions as a function of $\sigma_A$ (colorbar in C) and $Z$.
        (\textit{F}) Activator profiles (top) and fate patterns (bottom) as a function of $L/\ell_A$ with $L/\ell_I=2$ ($\sigma_A=0.05$).
        (\textit{G}) Activator profiles as a function of increasing $\sigma_A$. For each noise level, $L/\ell_A$ is chosen optimally, corresponding to stars in C.
        (\textit{H}) PI and CI contributions to the utility of the optimal patterns in G as a function of $\sigma_A$.
        (\textit{I}) Top 1\% utility parameter regions as a function of $L/\ell_A$ and $\sigma_A$. Stars correspond to the maximal utility solutions in C.
        (\textit{J}) Schematic of the network with an additional expander species $E$. 
        (\textit{K}) Example profiles of the activator (left, solid), expander (left, dashed), and expander-corrected profiles (right). 20 samples are shown in thin grey lines, with three examples highlighted in blue, green, red.
        (\textit{L}) Expander-corrected profiles $\tilde{A}$ for optimal parameter values of $L/\ell_A$ as a function of increasing extrinsic noise level $\sigma_A$.
        (\textit{M}) PI and CI contributions to the utility of the optimal expander-corrected profiles as a function of noise level.
        (\textit{N}) Top 1\% utility parameter regions of the expander network as a function of $L/\ell_A$ and $\sigma_A$.
        Throughout this example, we use a dimensionless parametrization, with the free parameters held fixed at $\alpha_A \beta_I / (\beta_A \alpha_I) = 0.5$ and $h=5$. We align activator profiles such that maxima are located at $x=L$, and use optimized thresholds. 
		 }
	\label{fig_ai}
\end{figure*}

\section{Example 3: Stochastic reaction-diffusion systems}
\label{sec_ex_rd}

A key developmental paradigm are cell fate patterns established by diffusible  morphogens~\cite{Green2015,Kicheva2023}. Here we study a minimal  model for this paradigm, i.e., a reaction-diffusion system with two chemical species -- an ``activator'' and an ``inhibitor''~\cite{Gierer1972,Segel1972} -- and search for its optimal self-organization parameters in the presence of noise. Activator-inhibitor patterning describes a number of developmental and regenerative systems~\cite{Kondo2010}, including palatal ridge formation~\cite{Economou2012}, digit patterning~\cite{Sheth2012}, and long-range axis patterning in planarian worms~\cite{Stueckemann2017}. 

Activator and inhibitor species are tracked in space and time by variables $A(x,t)$ and $I(x,t)$, respectively, which we collect into a two-component concentration vector $\boldsymbol{g}=(A,I)$ whose dynamics can be seen as a special case of \eqref{eq:dynsys}. The two species diffuse with coefficients $D_A$ and $D_I$, and interact with Hill-type reaction terms $f(A,I)=A^h/(A^h+I^h)$~\cite{Werner2015}:
\begin{align}
    \label{eq:ei_model}
    \partial_t A &= D_A \partial^2_xA + \alpha_A f(A,I) - \beta_A A + \sigma\xi(t)\nonumber \\
    \partial_t I &= D_I \partial^2_xI + \alpha_I f(A,I) - \beta_I I+ \sigma\xi(t),
\end{align}
where $\sigma$ denotes the intrinsic noise magnitude.
Motivated by the example of long-range axis patterning~\cite{Stueckemann2017}, we solve this system on a one-dimensional domain of length $L$ with reflecting boundary conditions. If the inhibitor diffuses much faster than the activator, spatial patterns emerge with shapes that depend on the ratios of the diffusive length scales of either species, $\ell_{A,I}=\sqrt{D_{A,I}/\beta_{A,I}}$, to the system size $L$. In our example, steady-state concentration patterns are subsequently interpreted by individual cells that threshold their local activator levels, $A(x_i,T\rightarrow\infty)$, to commit to one of the $Z$ possible cell fates. Ultimately, variability in the developmental ensemble is controlled by noise~(Fig.~\ref{fig_ai}A): intrinsic (here with fixed $\sigma=0.005$) as well as extrinsic, whose role we explore in depth in what follows. Here, we implement extrinsic noise as fluctuations in the activator production rate across replicates, $\alpha_A \sim \mathcal{N}(\bar{\alpha}_A,\sigma_A^2)$, and we treat its magnitude $\sigma_A$ as a parametric constraint. 

Utility depends non-trivially on two key dimensionless parameters, $L/\ell_A$ and $L/\ell_I$, revealing a broad utility optimum at intermediate activator diffusion (Fig.~\ref{fig_ai}B). To understand this optimum, we fix $L/\ell_I$ and vary $L/\ell_A$, taking the system through multiple regimes: at large activator diffusion (small $L/\ell_A$), there are no spontaneously generated patterns; this is followed by the onset of patterning in an intermediate $L/\ell_A$ regime with a single stable wavelength due to the finite system size (pattern monostability); while at small activator diffusion (large $L/\ell_A$), multiple wavelengths fit into the system, leading to pattern multistability~(Fig.~\ref{fig_ai}F). 

How do utility optima depend on the noise level? At small noise, monostability is preferred, with a broad flat maximum as a function of $L/\ell_A$~(Fig.~\ref{fig_ai}C). As the extrinsic noise is increased, the optimum becomes progressively narrower, and shifts towards larger values of $L/\ell_A$. This is because larger values of $L/\ell_A$ (i.e., smaller $D_A$) increase the pattern amplitude (Fig.~\ref{fig_ai}F), thereby increasing the effective signal-to-noise ratio that makes patterning more robust to fluctuations. At even larger extrinsic noise, the optimum transitions into the multistable regime: it is better, from the utility perspective, to further increase the pattern amplitude and maintain robust patterning in face of noise even though this splits the developmental ensemble into a mixture of at least two distinct ``body plans''~(Fig.~\ref{fig_ai}C, G).

This split is the reason why, at the onset of multistability, PI sharply drops and a large CI contribution to the utility emerges with increasing $L/\ell_A$ (Fig.~\ref{fig_ai}D). A similar effect is observed in systems with  $L/\ell_A$ optimized separately for each extrinsic noise level, as the noise increases (Fig.~\ref{fig_ai}H). In sum, when noise is sufficiently low for theoretically maximal utility values to be approached, maximizing utility corresponds to maximizing PI with vanishing CI, with an emergence of a single reproducible body plan, as we noted already in Section~\ref{sec_framework}; yet at higher noise, multistability of body plans cannot be avoided. Interestingly, if the number of possible cell fates $Z$ with which the system needs to be reliably patterned increases, noise represents a more severe constraint and the multistable patterns become optimal already at lower noise levels~(Fig.~\ref{fig_ai}E). Assuming that multistable body plans are not evolutionarily favored, we  wondered if additional regulatory mechanisms could implement error-correction to improve patterning outcomes even at high extrinsic noise.

To that end, we extended our basic model, \eqref{eq:ei_model}, by including an additional diffusible species, the ``expander'' $E$~\cite{Othmer1980,Ben-Zvi2010,Werner2015}, which feeds back onto the cell fate decision process~(Fig.~\ref{fig_ai}J). By considering multiple network topologies, we find that when fate decisions are made by thresholding a read-out species activated via an incoherent feed-forward loop~\cite{Alon2006,DeRonde2012} by $A$ while simultaneously inhibited by fast-diffusing $E$ (which is in turn also activated by $A$), extrinsic-noise-induced fluctuations are corrected (Fig.~\ref{fig_ai}K). When we optimize the parameters of this expander network (Supplementary Information), we are driven towards very fast diffusion for the expander species, allowing $E$ to essentially become a global ``sensor'' for the overall replicate-to-replicate amplitude variation of $A$ which can consequently be compensated for~(Fig.~\ref{fig_ai}L). With these optimal parameters, we find a reliable preference for the monostable regime of the underlying activator-inhibitor network even as extrinsic noise is increased, resulting in high utility values that lead to a unique and reliable body plan for which $U \approx$ PI  (Fig.~\ref{fig_ai}M,N). Taken together, our framework predicts both the topology and optimal parameter values for an ``expander'' module that can be added onto the classical activator-inhibitor system to drastically improve its patterning performance in the face of extrinsic variability.

\section{Discussion}
\label{sec_discussion}
\begin{figure*}
\centering
\includegraphics[width=0.7\textwidth]{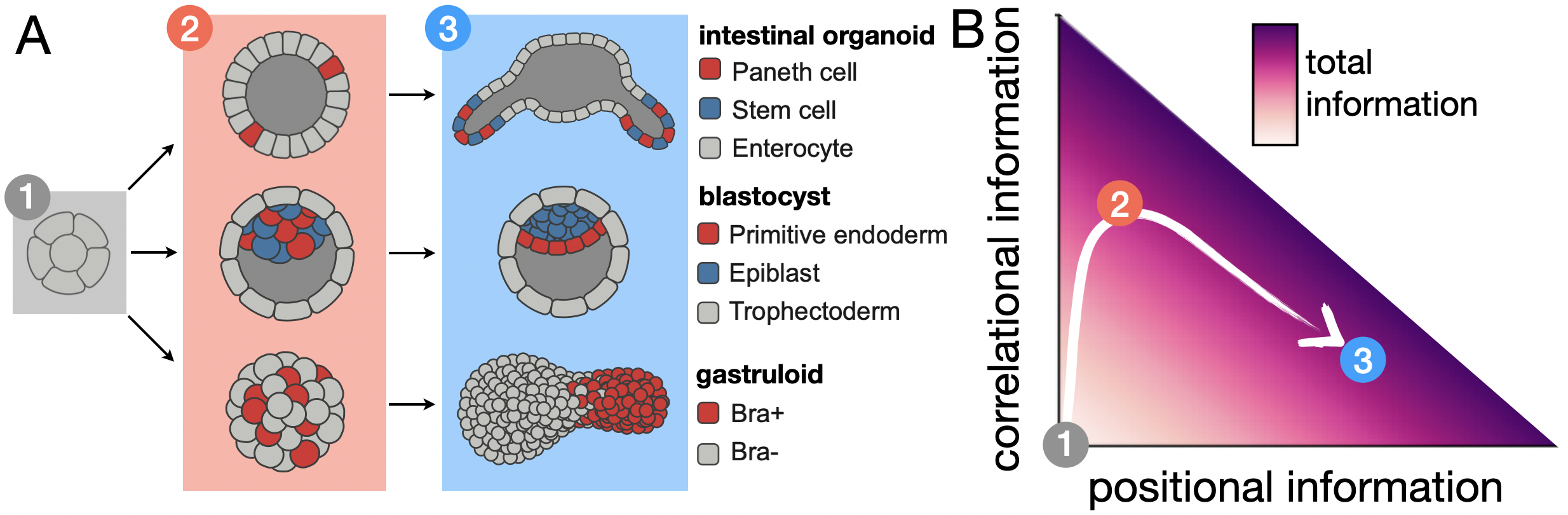}
    \caption{\textbf{Classification of self-organized systems in information space.} 
    (\textit{A}) Schematic of developmental stages of three self-organized developmental systems: \textit{in vitro} intestinal organoids (top), the early human embryo at the blastocyst stage (middle), and \textit{in vitro} 3D gastruloids.
    (\textit{B}) Hypothesized trajectory in the information plane: initial non-patterned stages have no information (1), intermediate stages give rise to correlational information (2), and final stages establish reproducible body plans with high positional information (3).
		 }
	\label{fig_discussion}
\end{figure*}

In this study, we introduced a mathematical framework that defines information content of self-organized spatial patterns. Maximization of this measure allowed us to identify optimal operating regimes of several paradigmatic developmental patterning systems in the presence of realistic sources of noise.  The results demonstrate how and why such noise can fundamentally alter the mechanisms required to generate reproducible patterns. Functionally, our framework allows us to rationalize the existence, and assess the importance, of individual mechanisms and modules that comprise biological patterning systems.

The utility measure, \eqref{eq_utility}, encapsulates a very generic trade-off: we hypothesize that an essential feature of self-organization in developmental systems is a simultaneous maximization of reproducibility and of cell type diversity. While we expect the reproducibility entropy term to be general, different alternatives to maximizing diversity are possible. For example, rather than being maximized, the cell type diversity could just need to be sufficient, as when two cell types should be generated reliably in known, yet unequal proportions~\cite{Saiz2016,Saiz2020}. Our framework accommodates this situation easily, by minimizing reproducibility entropy at fixed patterning entropy, formalizing the notion of robustness to noise~\cite{Barkai1997}. One can also contemplate more elaborate measures to be traded against the reproducibility, for example ``pattern complexity.'' To quantify complexity, a number of definitions based on algorithmic information theory have been introduced, including Kolmogorov complexity, effective complexity, logical depth, and complextropy~\cite{Lloyd2001}. While such complexity measures could be considered in future work as additional trade-off terms along the lines proposed previously~\cite{Gell-Mann1996}, one should bear in mind that these measures typically lack tractability (e.g., Kolmogorov complexity is uncomputable in most cases) or generality (e.g., assumptions must be made about what is  ``complex''). A less abstract set of alternatives would trade off reproducibility against a term that minimizes the divergence of the model-generated patterns from a desired (usually measured) pattern~\cite{Mlynarski2021}. This allows exploring trade-offs induced by biophysical constraints and maximizing reproducible patterning outcomes that are similar to the observed pattern~\cite{Francois2007a,Pezzotta2022,Sokolowski2023}. Another productive approach is to start with the positional information itself as the utility~\cite{Francois2010,Sokolowski2023} to be maximized. Here, our results provide the theoretical justification for this proxy, which should be relevant (i) in the low noise regime; and (ii), in the stages of patterning after any spontaneous symmetry breaking has already occurred. 


Information content of an ensemble of patterns can be decomposed into two interpretable contributions. Positional information (PI) measures the local spatial ordering that directly reflects the specification of a reproducible body plan. Correlational information (CI) quantifies the amount of non-local statistical structure that increases developmental reproducibility but does not directly imply a tight correspondence between cell types and positions. Applied to example systems we consider, this decomposition suggests that, interestingly, development might undergo  phases that are dominated by CI, followed by transformation to states with high PI (Fig.~\ref{fig_discussion}). Firstly, intestinal organoids break symmetry through lateral inhibition, thereby selecting a positionally random, but statistically correlated, subset of cells to differentiate into Paneth cells, which then secrete morphogen signals to build a positionally ordered crypt pattern~\cite{Serra2019}. Secondly, the inner cell mass compartment of the early human embryo self-organizes precise proportions of two cell types (primitive endoderm and epiblast)~\cite{Saiz2016,Saiz2020}, followed by sorting through differential adhesion into a positional pattern~\cite{Yanagida2022}. Finally, 3D gastruloids initially establish salt-and-pepper patterns of Brachyury positive and negative cells~\cite{Hennessy2023}, while later stages exhibit remarkably precise positional patterns~\cite{Merle2023} (Fig.~\ref{fig_discussion}A); a similar transformation has also been observed in neural tube organoids~\cite{Krammer2023}. All these systems appear to initially generate CI through symmetry breaking, which is then redistributed into PI -- our suggestion is to visualize and analyze these processes as trajectories in the information plane (Fig.~\ref{fig_discussion}B). Other, not fully self-organized, developmental systems can also be understood in terms of the same framework. For instance, 2D stem cell assemblies self-organize concentric patterns of cell fates~\cite{Warmflash2014,Chhabra2019a,Lehr2023}, albeit based on an initial ``boundary condition'' of receptor localization~\cite{Etoc2016}. Furthermore, self-organized lateral inhibition signaling in \textit{Drosophila} patterning operates on a long wavelength ``pre-pattern''~\cite{Schweisguth2019}. Understanding how all these systems set up and transform information as the developmental processes unfold may provide a unifying classification scheme for patterning mechanisms and suggest generic routes towards self-organized patterns and body plans with high PI.

Self-organized fate patterning proceeds through a sequence of steps, from noisy initial conditions, to patterns of signaling activity, to cell fate commitment. Self-organization through cell-cell communication has been described by physical models ranging from reaction-diffusion systems to contact-based interactions~\cite{Green2015,Bailles2022a,Kicheva2023}. Fate specification at the single-cell level has been described by intra-cellular gene regulatory networks and landscape-based dynamical systems models~\cite{Corson2012,Rand2021}. An attractive view is that both steps can be understood in the language of dynamical systems, which provides strong constraints on the type of bifurcation motifs that comprise these systems~\cite{Rand2021}. We build on this integrative view to consider the entire developmental sequence, followed by the utility-based evaluation of its final outcome. This allows us to ask why, given a certain structure and magnitude of various noise sources, particular regulatory motifs should be favored. It also allows us to ask how the established signals would be best compressed into discrete cell fates~\cite{Bauer2021}. As demonstrated by the  mono- vs bistable lateral inhibition signaling example (Fig.~\ref{fig_LIS_bistable}), different motifs indeed can generate the same amount of information in the absence of noise, but exhibit divergent performance at larger noise magnitudes. Such optimization in the presence of noise would be interesting to incorporate into functional large-scale screens of computational patterning models, such as multi-species Turing systems~\cite{Marcon2016a,Scholes2019}, which have thus far been limited to deterministic regimes. Furthermore, our example of cell type proportioning shows how under finite-time constraints, addition of noise can actually increase utility, which is in line with a recently suggested noise-dependent mechanism in spatial patterning of bacterial biofilms~\cite{Nadezhdin2020}. More generally, our results reinforce the notion that noise is not just a ``small correction'' to patterning (even if it appears as such in evolved systems that we experimentally study), but -- in the broad space of evolutionary possibilities -- should be simultaneously considered both as a deleterious force to be kept in check as well as a source of variability to be harnessed during early development.

Beyond optimization, our framework lays the foundation for estimating the total information content of experimentally observed patterns. Specifically, we  provide a generalization of positional information measure to the total information content of a pattern, which also includes CI. In the context of morphogen patterns that are informative about position, CI lower-bounds the excess information that may be gained by making use of spatial correlations. This suggests a possible generalization of local~\cite{Petkova2019} to non-local decoding. Indeed, recent work demonstrates that additional information is contained in spatial correlations of the pair rule stripes in \textit{Drosophila}~\cite{McGough2023}. Here we entertained the minimal idea that permitted converting CI into PI through a fast-diffusing expander species~\cite{Othmer1980,Ben-Zvi2010,Werner2015} integrated as an incoherent feed-forward loop~\cite{Alon2006,DeRonde2012,Francois2013}. The circuit that can compensate for the effects of extrinsic noise represents a form of spatial buffering, in analogy with temporal buffering in circadian clocks, where a long lived chemical builds up  robustness against temperature changes~\cite{Nakajima2005}. Similar principles could apply to sensing mechanisms in a local neighborhood, as was recently suggested in the context of epiblast patterning in a ``neighbourhood watch model''~\cite{Lee2022a}.

While our primary focus was to formalize the notion of information establishment in self-organized systems and illustrating it on a broad range of well-understood toy models, the application of this framework to large-scale high-dimensional datasets necessitates further development of sophisticated computational, approximative, and inference schemes. In future work, this could allow estimation of information content from experimental data sets of spatio-temporal signaling and gene expression, such as multiplex staining~\cite{Gut2018}, high-throughput live imaging~\cite{DeMedeiros2022}, single-cell sequencing~\cite{Casey2023} or spatial transcriptomics~\cite{Crosetto2015}. Linking our optimality framework to rigorous statistical inference and hypothesis testing via a recently proposed Bayesian framework~\cite{Mlynarski2021} would allow quantitative, data-driven inference of the patterning performance from experimental data.

\section{Acknowledgements}
We thank Wiktor M{\l}ynarski, Juraj Majek, Michal Hled\'ik, Fridtjof Brauns, Nikolas Claussen, Benjamin Zoller, Erwin Frey, Thomas Gregor, and Edouard Hannezo for inspiring discussions. D.B.B. was supported by the NOMIS foundation as a NOMIS Fellow and by an EMBO Postdoctoral Fellowship (ALTF 343-2022). This research was performed in part at the Aspen Center for Physics, which is supported by NSF grant No. PHY-1607611, and KITP Santa Barbara, supported by NSF Grant No. PHY-1748958 and the Gordon and Betty Moore Foundation Grant No. 2919.02.

%



\section*{Supplementary material}

# Supplementary Information:
## Information content and optimization of self-organized developmental systems

David B. Brückner and Gašper Tkačik

## Contents

## 1 Derivation of information-theoretic quantities

Here, we provide a more detailed derivation of the key relationships of the information-theoretic quantities introduced in the main text. Specifically, we derive the following set of relationships:

$$U = \frac{1}{N} D_{\mathrm{KL}}[P(\vec{z})||Q(\vec{z})] = \mathrm{PI} + \mathrm{CI} = S_{\mathrm{pat}} - S_{\mathrm{rep}} \tag{S1}$$

Here, as defined in the main text, the distribution $Q(\vec{z})$ is the is the maximum entropy distribution given the observed pooled distribution

$$Q(\vec{z}) = \prod_{i=1}^{N} P_z(z_i). \tag{S2}$$

The pooled distribution $P_z(z)$ is defined as the probability that a randomly chosen cell has fate $z$, and is therefore given by the average of $P(z_i = z)$ across positions:

$$P_z(z) = \frac{1}{N} \sum_{i=1}^{N} P(z_i = z) = \frac{1}{N} \sum_{i=1}^{N} \sum_{z_i=1}^{Z} P(\vec{z}) \delta(z_i, z) \tag{S3}$$

since

$$P(z_i = z) = \sum_{z_i=1}^{Z} P_i(z_i) \delta(z_i, z) = \sum_{z_i=1}^{Z} P(\vec{z}) \delta(z_i, z). \tag{S4}$$

To derive the relationships in Eq. (S1), we start the derivation with the KL divergence, show that this decomposes into PI and CI, and then show that this is given by the

1

difference of patterning and reproducibility entropies. First, we note that

$$U = \frac{1}{N} D_{\text{KL}}[P(\vec{z})||Q(\vec{z})] = \frac{1}{N} \sum_{\vec{z}} P(\vec{z}) \log_2\left(\frac{P(\vec{z})}{Q(\vec{z})}\right) \tag{S5}$$

$$= -S_{\text{rep}} - \frac{1}{N} \sum_{\vec{z}} P(\vec{z}) \log_2(Q(\vec{z})) \tag{S6}$$

$$= -S_{\text{rep}} - \frac{1}{N} \sum_{\vec{z}} P(\vec{z}) \sum_{i=1}^{N} \log_2(P_z(z_i)) \tag{S7}$$

$$= -S_{\text{rep}} - \frac{1}{N} \sum_{i=1}^{N} \sum_{z_i=1}^{Z} P_i(z_i) \log_2(P_z(z_i)) \tag{S8}$$

where we have used the definition of the marginal distribution $P_i(z_i)$ in the last step:

$$P_i(z_i) = \sum_{z_1=1}^{Z} \cdots \sum_{z_{i-1}=1}^{Z} \sum_{z_{i+1}=1}^{Z} \cdots \sum_{z_N=1}^{Z} P(\vec{z}) \tag{S9}$$

To simplify the second term in Eq. (S8), we write

$$U = -S_{\text{rep}} - \frac{1}{N} \sum_{i=1}^{N} \sum_{z=1}^{Z} P_i(z_i) \log_2\left(P_z(z_i) \frac{P_i(z_i)}{P_i(z_i)}\right) \tag{S10}$$

$$= -S_{\text{rep}} + \frac{1}{N} \sum_{i=1}^{N} S[P_i(z_i)] + \frac{1}{N} \sum_{i=1}^{N} D_{KL}[P_i(z_i)||P_z(z_i)] \tag{S11}$$

Here, we identify the second term as the entropy of a correlation-free (cf) joint distribution, which is simply given by the product of the marginals:

$$S_{\text{cf}} = \frac{1}{N} S\left[\prod_{i=1}^{N} P_i(z_i)\right] \equiv S[P_{\text{cf}}(\vec{z})] \tag{S12}$$

Then, considering the first two terms together,

$$S_{\text{cf}} - S_{\text{rep}} = -\frac{1}{N} S[P(\vec{z})] - \frac{1}{N} \sum_{\vec{z}} P(\vec{z}) \log_2(P_{\text{cf}}(\vec{z})) \tag{S13}$$

$$= \frac{1}{N} D_{KL}[P(\vec{z})||P_{\text{cf}}(\vec{z})] \tag{S14}$$

where we used Eq. (S9) in the first equality. This shows that the first two terms in Eq. (S11) are the reduction in entropy due to the presence of correlations in $P(\vec{z})$, which can be interpreted as the information contained in these correlations. We therefore call this the *correlational information* CI:

$$\text{CI} = S_{\text{cf}} - S_{\text{rep}}. \tag{S15}$$

To understand the third term in Eq. (S11), we consider that

$$\frac{1}{N}\sum_{i=1}^{N} D_{KL}[P_i(z_i)||P_z(z)] = \frac{1}{N}\sum_{i=1}^{N}\sum_{z=1}^{Z} P_i(z_i)\log_2\left(\frac{P_i(z_i)}{P_z(z)}\frac{P_{\text{index}}(i)}{P_{\text{index}}(i)}\right) \quad \text{(S16)}$$

$$= \sum_{i=1}^{N}\sum_{z=1}^{Z} P(z,i)\log_2\left(\frac{P(z,i)}{P_z(z)P_{\text{index}}(i)}\right) \quad \text{(S17)}$$

$$= I(z;i) = \text{PI} \quad \text{(S18)}$$

where we have used Bayes' rule $P(z,i) = P_i(z_i)P_{\text{index}}(i)$, we used the fact that cell indices are uniformly distributed by definition, i.e. $P_{\text{index}}(i) = 1/N$. $I(z;i)$ is the mutual information between cell fate $z$ and cell position $i$, which is the *positional information* of the pattern PI [1–3]. Putting everything together, we have

$$U = \text{PI} + \text{CI} \quad \text{(S19)}$$

Thus, the utility function measures the total information in the pattern, which is the sum of positional and correlational information.

To demonstrate the last equality in Eq. (S1), i.e. relating the KL divergence to the difference of patterning and reproducibility entropies, we note that positional information is given by a difference of entropies,

$$\text{PI} = S[P_z(z)] - \sum_{i=1}^{N} P_{\text{index}}(i)S[P_i(z_i)] \quad \text{(S20)}$$

$$= S[P_z(z)] - \frac{1}{N}\sum_{i=1}^{N} S[P_i(z_i)] \quad \text{(S21)}$$

$$= S_{\text{pat}} - S_{\text{cf}} \quad \text{(S22)}$$

where in the last step, we again assumed uniform distribution of cell positions. Inserting Eqs. (S15) and (S22) into Eq. (S19), we then obtain

$$U = S_{\text{pat}} - S_{\text{rep}}. \quad \text{(S23)}$$

## 2 Lateral inhibition signaling

As a minimal model of lateral inhibition signaling (LIS), we use a generic dynamical system of cell-cell signaling,

$$\frac{dg_i}{dt} = F(g_i, s_i) - g_i + \sigma\xi(t), \quad \text{(S24)}$$

with all variables as defined in the main text, and

$$F(g_i, s_i) = \frac{1}{1 + \exp[-f(g_i, s_i)]}. \quad \text{(S25)}$$

3

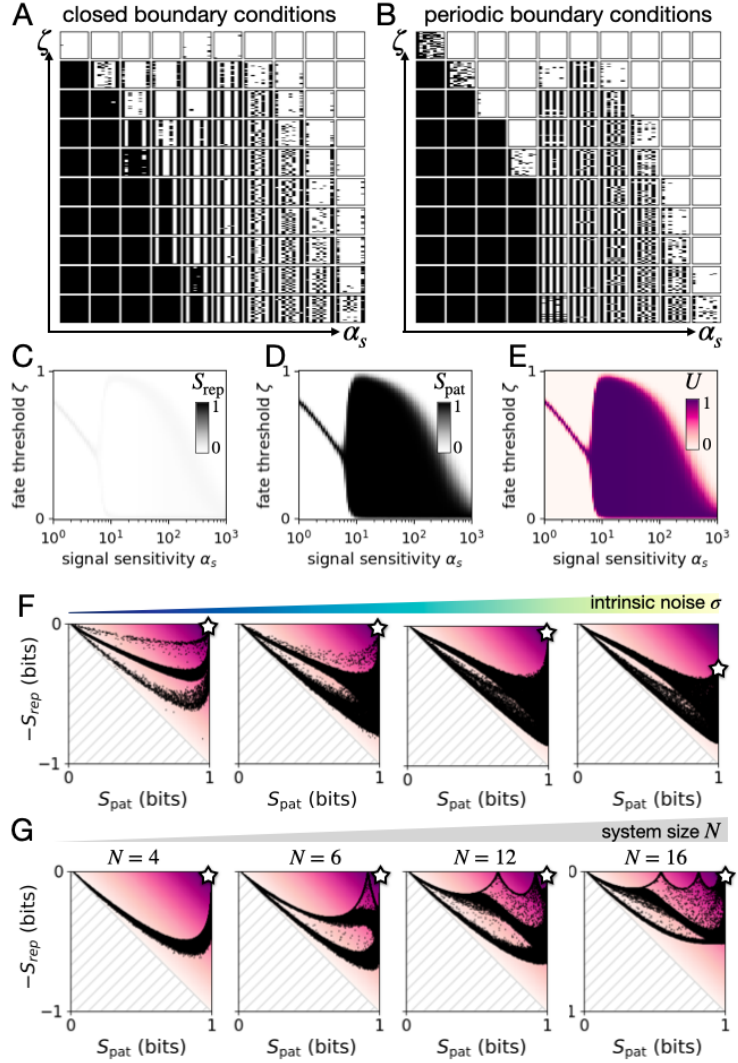

Figure S1: **Supplementeray results on lateral inhibition signaling without self-activation.** (*A,B*) Developmental ensembles as a function of parameters $\alpha_s$ and $\zeta$ for closed and periodic boundary conditions, respectively. (*C,D,E*) Reproducibility entropy, patterning entropy, and utility, respectively, as a function of $\alpha_s$ and $\zeta$, with fixed $\sigma = 0.01$, for periodic boundary conditions. (*F*) Visualization of patterning outcomes with periodic boundary conditions in entropy planes for four increasing intrinsic noise levels ($\sigma = \{0.001, 0.01, 0.05, 0.1\}$). Each of the $10^5$ dots corresponds to a developmental ensemble defined by a random draw of its parameters $\theta = \{\alpha_s, \zeta\}$. (*G*) Visualization of patterning outcomes as a function of increasing cell number with closed boundary conditions.

We use this formulation of $F(g_i, s_i)$ as it comprises a large class of models that have been used in the literature to model gene regulation and cell signaling. We highlight these connections below.

First, a previous model of delta-notch signaling that included a bistability used the function [4]

$$F(g_i, s_i) = \frac{1 + \tanh(4[g_i - s_i])}{2}, \tag{S26}$$

4

which is equivalent to our formulation using

$$f(g_i, s_i) = \alpha_g g_i - \alpha_s s_i \tag{S27}$$

with $\alpha_g = \alpha_s = 8$.

Second, using

$$f(g_i, s_i) = -h_s \log(s/K_s), \tag{S28}$$

we obtain Hill function familiar from chemical kinetics, with Hill coefficient $h_s$ and parameter $K_s$,

$$F(g_i, s_i) = \frac{1}{1 + (s/K_s)^{h_s}}. \tag{S29}$$

Finally, using

$$f(g_i, s_i) = -h_s \ln\left[\frac{1 + s_i/K_s}{1 + 1/(2K_s)}\right] + h_g \ln\left[\frac{1 + g_i/K_g}{1 + 1/(2K_g)}\right] \tag{S30}$$

we obtain the Monod-Wyman-Changeux functional form first used to describe allosteric transitions in proteins and later generalized to two-state transitions in regulatory processes like gene regulation [5]. Note that both Eqs. (S28) and (S30) couple to inputs using two parameters per input (a "steepness-like" parameter $h$ and a "threshold-like" parameter $K$), so they are more expressive than the simple linear coupling for $f$ we use for illustration purposes in the main paper; nevertheless, both models reduce to the linear $f$ when $s \ll K$.

Parameter sweeps over Eqs. (S28) and (S30) show qualitatively similar behaviour of reproducibility entropy, patterning entropy, and utility as displayed in main text Fig. 3 and 4.

Here we show supplementary results to main text Fig. 3, i.e. using the regulatory function without self-activation, $f(s_i) = -\alpha_s s_i$. First, we show the developmental ensembles corresponding to the main text Fig. 3B-D (Fig. S1A), and those predicted by an equivalent parameter sweep with periodic boundary conditions (Fig. S1B). To analyze the setting with periodic boundary conditions from the standpoint of our framework, Figs. S1C-E show the corresponding reproducibility entropy, patterning entropy, and utility; and Fig. S1F the patterning outcomes in the entropy plane as a function of increasing noise. Finally, Fig. S1G shows the patterning outcomes in the case with closed boundary conditions for varying cell number $N$, showing how the point cloud shape changes with the system size.

To explore the role of an additional self-activation, we use the regulatory input function $f(s_i) = -\alpha_s s_i + \alpha_g g_i$. As a function of $\alpha_s$ and $\alpha_g$, the nullclines of Eq. (S24) exhibit an intricate dependence with bistability emerging for a range of values (Fig. S2A). In the main text, we explored this space along the diagonal, using a single parameter $\alpha = \alpha_s = \alpha_g$ (blue curves in Fig. S2A). Here we show the dependence of utility on the full $(\alpha_s, \alpha_g)$-space (Fig. S2B-E). Furthermore, we show the dependence of the utility landscape on the magnitude of intrinsic noise (Fig. S2F).

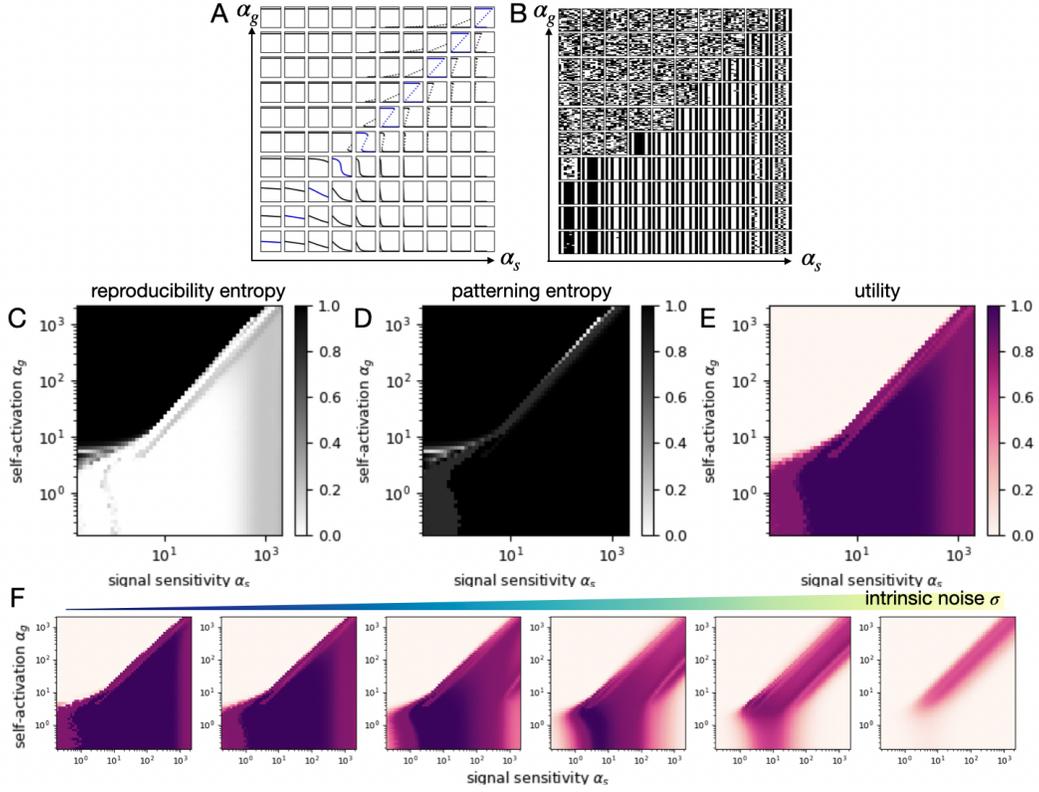

Figure S2: **Lateral inhibition signaling with self-activation.** (*A*) Nullclines of Eq. (S24) as a function of $(\alpha_s, \alpha_g)$. Solid and dotted lines correspond to stable and unstable fixed points, respectively. Parameter combinations used in the main text along the diagonal $\alpha = \alpha_s = \alpha_g$ are colored blue. (*B*) Developmental ensembles as a function of parameters $\alpha_s$ and $\alpha_g$ for closed boundary conditions, with optimized thresholds. (*C,D,E*) Reproducibility entropy, patterning entropy, and utility, respectively, as a function of $\alpha_s$ and $\alpha_g$, with fixed $\sigma = 0.01$. (*F*) Utility landscape as a function of increasing intrinsic noise.

# 3 Cell type proportioning and sorting

## 3.1 Maximal CI generated by proportioning

In cell type proportioning, we consider mechanisms with no local coupling, meaning cells only sense a global signal which instructs their fate decision making. In this case, all information is correlational information, as there can be no correlations with position. We can calculate the upper bound of the correlational information by considering how many permutations of a given proportioning there are. Specifically, suppose in each system with $N$ cells, there are $\{n_1, ..., n_k, ..., n_Z\}$ cells with each of the cell fates $1...Z$; $\sum_{k=1}^{Z} n_k = N$. Then, the number of possible permutations are

$$W = \frac{N!}{n_1!...n_k!...n_Z!} \tag{S31}$$

6

Since we are looking for the ensemble with maximal correlational information, we are considering proportionings with maximal patterning entropy $S_{\text{pat}} = \log_2 Z$, corresponding to the case where all $Z$ fates occur with equal probability. For simplicity, we consider the case where $N$ is divisible by $Z$, meaning that in each system, equal numbers of each fate can be realized. Then, $n_k = N/Z$ for all $k$, and therefore

$$W = \frac{N!}{[(N/Z)!]^Z} \tag{S32}$$

Each of these $W$ states occurs with probability $1/W$, and thus

$$S_{\text{rep}} = \frac{1}{N} \log_2 W \tag{S33}$$

which gives a maximal correlational information

$$\text{CI}_{\text{max}} = \log_2 Z - \frac{1}{N} \log_2 W \tag{S34}$$

This is the upper bound of proportioning-generated CI shown in Fig. 5 of the main text. This upper bound decreases with increasing cell number $N$, since the reduction in entropy due to perfect proportioning decreases as system size increases (Fig. S3A, B). In other words, correlations that generate perfect proportioning generate a correction to CI that is sub-extensive, hinting that these mechanisms may be more important at very low $N$ (and thus at initial stages of development).

## 3.2 Minimum PI generated by sorting

If a system with proportions of different cell types is sorted according to cell fate, PI is generated. When the initial proportioning is perfect, meaning $U = \text{CI}_{\text{max}}$, then the final utility of the sorted system is $U = \text{PI}_{\text{max}} = \log_2 Z$ bits (assuming perfect, noise-free sorting). However, even if the initial proportioned system is generated by completely random, uncorrelated cell fate decisions, such a sorting process will generate PI. Here, we show how to calculate this lower bound of sorting-generated PI.

For this, we assume that the proportioning mechanism generates $Z = 2$ cell fates with equal probability,

$$P(z = 1) = P(z = 2) = 0.5, \tag{S35}$$

but there is no communication between cells, i.e. cell decisions are a Bernoulli process. Therefore, $P_z(z)$ is a uniform distribution, and $S_{\text{pat}} = 1$ bit. The developmental ensemble is then given by the binomial distribution of the number of cells with either of the two fates, which is the maximum entropy distribution, $P(\vec{z}) = Q(\vec{z})$. Thus, $S_{\text{rep}} = 1$ bit, and $U = 0$.

Once this system is sorted, the marginal distributions are given by the cumulative dis-

7

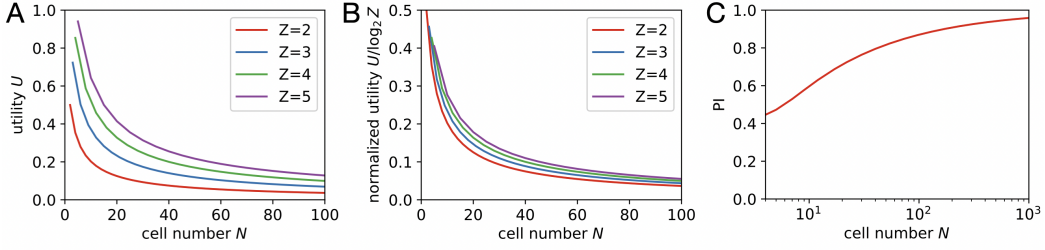

Figure S3: **Dependence of $CI_{max}$ and $PI_{min}$ on $N$ and $Z$.** (*A*) In case of perfect proportioning, $U_{max} = CI_{max}$, shown as a function of $N$ and $Z$ (colors), computed using Eq. (S34). (*B*) Same quantity as in A but normalized for the number of cell fates $Z$, i.e., $U_{max}/\log_2 Z$. (*C*) Minimum PI generated through noise-free sorting of a binomial ensemble, $PI_{min}$, as a function of cell number $N$ for $Z = 2$.

tributions of the binomial distribution:

$$S[P_i(z_i)] = -\sum_{z=1}^{Z} P(z_i = z) \, \log_2 P(z_i = z) \tag{S36}$$

$$= -P(N_1 \geq i) \, \log_2 P(N_1 \geq i) - P(N_2 \geq N - i + 1) \, \log_2 P(N_2 \geq N - i + 1) \tag{S37}$$

where $P(N_z \geq x)$ is the probability that $N_z$, the number of cells with fate $z$ in a single replicate, is greater or equal to $x$. This probability is given by the cumulative binomial distribution.

Based on these marginal distributions, we can then calculate the PI of a sorted binomial ensemble:

$$PI_{min} = 1 - \frac{1}{N} \sum_i S[P_i(z_i)]. \tag{S38}$$

This gives the minimum PI achievable in absence of any proportioning process, based on a binomial ensemble with equal fate probabilities. Note that as $N \to \infty$, $PI_{min} \to 1$ bit, as the contribution of the "fuzziness" of the boundaries between the cell types becomes negligible (Fig. S3C). Accordingly, the initial correlational information of the unsorted pattern is more important for the final PI generated through sorting in small systems, as we noted above.

## 4 Stochastic reaction-diffusion dynamics

### 4.1 Activator-Inhibitor system

Here, we provide additional detail on the analysis of the activator-inhibitor system. Specifically, we show behaviours in the 2D parameter space of activator and inhibitor length scales.

8

As described in the main text, we explore a minimal activator-inhibitor system described by the equations [6]

$$\partial_t A = D_A \partial_x^2 A + \alpha_A f(A, I) - \beta_A A + \sigma \xi(t)$$
$$\partial_t I = D_I \partial_x^2 I + \alpha_I f(A, I) - \beta_I I + \sigma \xi(t), \tag{S39}$$

with $f(A, I) = A^h/(A^h + I^h)$. Throughout this analysis, we start with noisy initial conditions of the activator $A(x, t = 0)$, implemented as uniform random numbers on the interval [0,0.1]. Here, we fixed $h = 5$ and $\alpha_A \beta_I / (\beta_A \alpha_I) = 0.5$ throughout, while varying the key length scales of the system,

$$\ell_A = \sqrt{D_A/\beta_A} \quad \text{and} \quad \ell_I = \sqrt{D_I/\beta_I}. \tag{S40}$$

This model exhibits a hierarchy of steady-state patterns, which are characterized by the number of contiguous source regions (defined as those regions in which $A > I$) $m$, and the number of source regions at the system boundary $n$ [6]. As a function of the length-scale parameters (Eq. (S40)), different $(m, n)$-patterns become stable, and lead to different relative frequency of these patterns as a function of parameters (Fig. S4A, B). This leads to absence, mono- or multi-stability of fate patterns as a function of parameters (Fig. S4C). In the zero- or low-noise limit, reproducibility and patterning entropies, and utility are fully determined by these relative fractions (Fig. S4D-F). As described in the main text, utility exhibits a pronounced maximum at intermediate activator and high inhibitor diffusion (corresponding to intermediate $L/\ell_A$ and low $L/\ell_I$; Fig. S4F). This overlaps with the monostable region, in which 100% of profiles correspond to (1,1)-patterns (Fig. S4G).

We next explore the behaviour of utility as extrinsic noise increases. Here, we implement extrinsic noise as fluctuations in the production rate of the activator $\alpha_A \sim \mathcal{N}(\bar{\alpha}_A, \sigma_A^2)$. In this case, not only the mono- vs multi-stability of the patterns (Fig. S4G), but also the dynamic range of the activator profiles becomes relevant, defined as $A_{\max} - A_{\min}$. This dynamic range increases with decreasing activator diffusion (i.e., increasing $L/\ell_A$) for all values of inhibitor diffusion (Fig. S4H). This suggests a generic trade-off: decreasing $L/\ell_A$ leads to monostability of patterns, but increasing $L/\ell_A$ increases dynamic range.

### 4.2 Noise-correcting expander network

To explore how CI can be transformed into PI in the reaction-diffusion example, we consider a minimal noise-correcting network that includes and expander species $E$. Here, we provide additional details on the implementation of this model.

For this, we build on the activator-inhibitor example, with the aim of finding a second subnetwork that takes the activator profile as an input and corrects the noise variations in it. In principle, reaction dynamics in this second subnetwork would take place simultaneously with the pattern formation in the first subnetwork. However, for computational simplicity, we here consider the case where the final activator profile

9

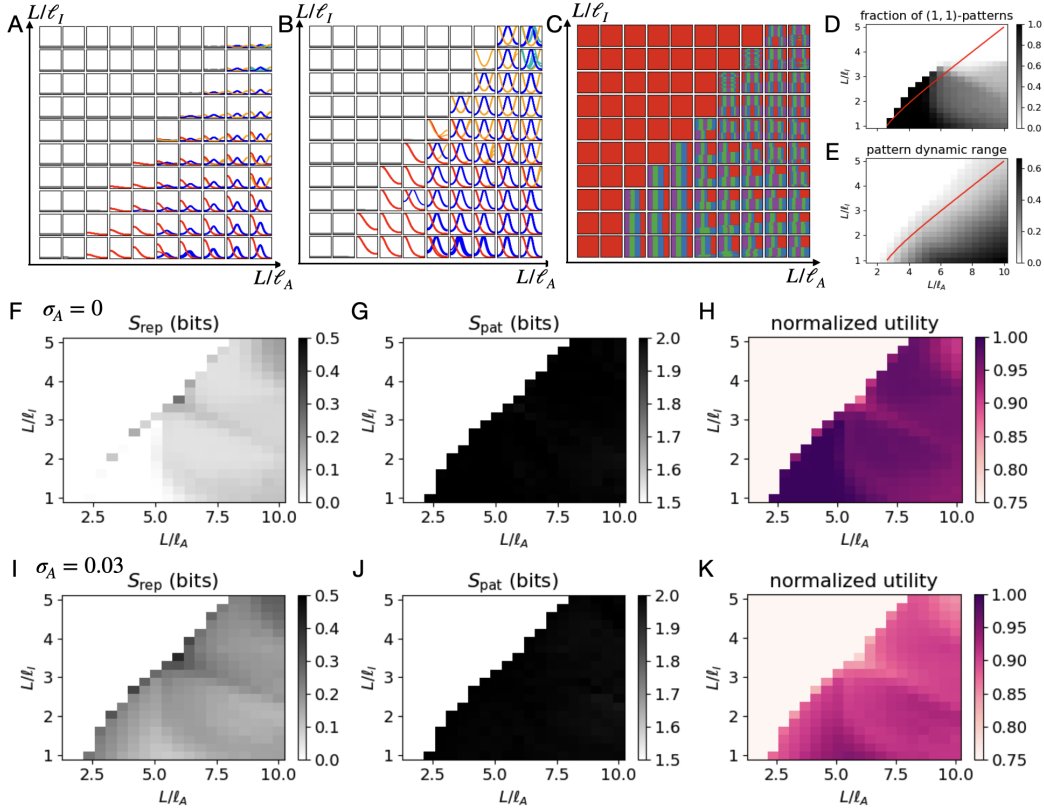

Figure S4: **Reaction-diffusion patterns as a function of parameters.** (A) Activator patterns $A(x, T)$ as a function of $L/\ell_A$ and $L/\ell_I$ (both parameters are varied on the same range as on the subsequent panels) for $\sigma_A = 0$. Color indicates the pattern type (m,n), see text: red (1,1), blue (1,0), orange (2,2), cyan (2,1). (B) Normalized activator patterns $A(x, T)/A_{\max}$. (C) Developmental ensembles based on activator patterns in A with optimized thresholds for $Z = 4$ fates. (D) Fraction of (1,1)-patterns as a function of $L/\ell_A$ and $L/\ell_I$. (E) Dynamic range of the activator pattern, defined as $A_{\max} - A_{\min}$. (F,G,H) Reproducibility entropy, patterning entropy, and normalized utility $U/\log_2 Z$ as a function of $L/\ell_A$ and $L/\ell_I$ for $\sigma_A = 0$. H corresponds to the plot in the main text. (I,J,K) Reproducibility entropy, patterning entropy, and normalized utility $U/\log_2 Z$ as a function of $L/\ell_A$ and $L/\ell_I$ for $\sigma_A = 0.03$.

$A(x) = A(x, t = T)$ is taken as an input. Note that parallel evolution could play an important role in the case of intrinsic noise, as it may allow noise-averaging, which we don't consider here. We find that a subnetwork that leads to noise corrections is the following: a non-diffusible species $\tilde{A}$ is activated by $A$, but inhibited by a diffusible expander species $E$, which in turn is activated by $A$. Specifically,

$$\partial_t \tilde{A}(x, t) = \alpha_C A(x) - \beta_C E(x, t) \tilde{A}(x, t) \tag{S41}$$

$$\partial_t E(x, t) = D_E \partial_x^2 E + \alpha_E A(x) - \beta_E E(x, t). \tag{S42}$$

To understand how this network leads to noise correction, we consider the limit of large expander diffusion $D_E$. First, we assume that the expander diffuses sufficiently fast that we can consider Eq. (S42) as an ODE and solve for an approximately flat steady-state expander profile $E(x) \approx E_0$. This constant value is is determined by the average

10

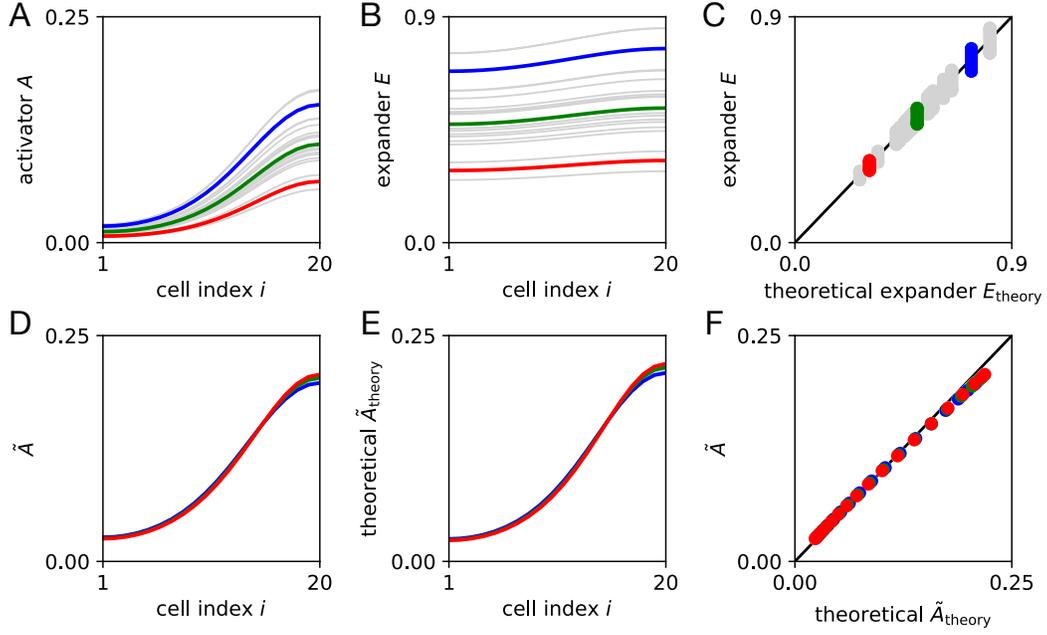

Figure S5: **Expander Mechanism.** (*A*) Activator patterns $A(x, T)$ for $\sigma_A = 0.05$. A collection of patterns is shown in light grey, with three patterns highlighted in red, green, blue throughout the figure. (*B*) Corresponding expander patterns $E(x, T)$. (*C*) Correlation between the actual and predicted expander profile based on Eq. (S43). (*D*) Pattern of the downstream species $\tilde{A}(x, T)$. (*E*) Predicted pattern of the downstream species $\tilde{A}(x, T)$ using Eq. (S44). (*F*) Correlation between actual and predicted pattern of the downstream species $\tilde{A}(x, T)$.

amplitude of the activator:

$$E_0 \approx \frac{\alpha_E}{\beta_E} \langle A(x) \rangle_x. \tag{S43}$$

Secondly, from Eq. (S41), we solve for steady-state, obtaining

$$\tilde{A}(x) = \frac{\alpha_C}{\beta_C} \frac{A(x)}{E(x)} \approx \frac{\alpha_C \beta_E}{\beta_C \alpha_E} \frac{A(x)}{\langle A(x) \rangle_x}. \tag{S44}$$

Therefore, $\tilde{A}(x)$ is a normalized version of $A(x)$, removing the extrinsic noise. In the limit of large $D_E$, we confirm these approximations numerically, and find that the profiles collapse with high accuracy (Fig. S5).

11

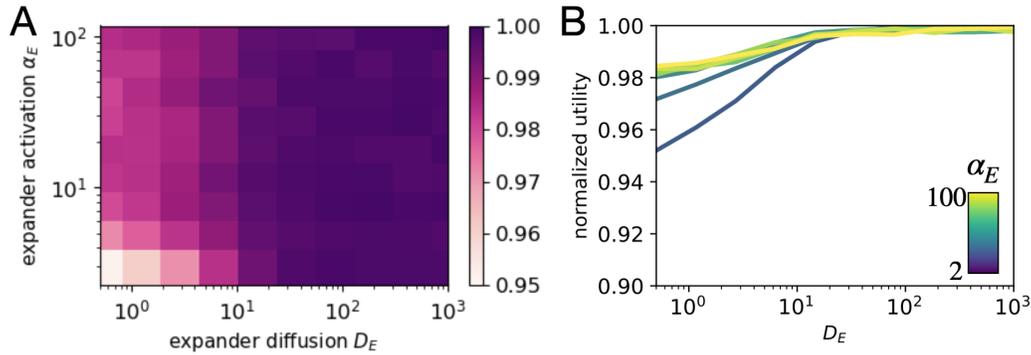

Figure S6: **Expander Mechanism.** (*A*) Normalized utility $U/\log_2 Z$ as a function of the expander diffusion $D_E$ and the expander activation rate $\alpha_E$. (*B*) Normalized utility $U/\log_2 Z$ as a function of the expander diffusion $D_E$ for varying expander activation rate $\alpha_E$ (colors).



\end{document}